\documentclass[review]{elsarticle}
\graphicspath{ {./figures/} }
\usepackage{hyperref}
\usepackage{float}
\usepackage{verbatim} 
\usepackage{apalike}
\restylefloat{figure}
\restylefloat{table}
\usepackage{pythonhighlight}
\usepackage{balance} 
\usepackage{tabularx}
\usepackage{makecell}
\usepackage{tocbibind}
\usepackage{verbatim}
\usepackage{longtable}
\usepackage{multicol}
\usepackage{multirow} 
\usepackage{amsmath}
\usepackage{tabularx} %
\usepackage[noend]{algpseudocode}
\usepackage{rotating}
\usepackage{caption}
\usepackage{threeparttable}
\usepackage{amssymb}
\usepackage[a4paper, margin=1.2 in]{geometry}
\usepackage{caption}
\usepackage{amsmath}
\usepackage{cleveref}  
\usepackage[T1]{fontenc}
\usepackage{algpseudocode}
\usepackage{xcolor}
\usepackage{algorithm2e}
\usepackage{geometry}
\usepackage{longtable}
\usepackage{array}

\usepackage{xspace}
\usepackage{booktabs}
\usepackage{graphicx}
\usepackage{subcaption}
\usepackage[labelformat=simple]{caption}

\journal{
}

\biboptions{authoryear}

\begin{document}
\begin{frontmatter}

\title{Collaborating in a competitive world: Heterogeneous Multi-Agent Decision Making in Symbiotic Supply Chain Environments
}

\author[label1,label2]{Wan Wang \corref{cor1}}
\ead{ww1a23@soton.ac.uk}

\author[label1]{Haiyan Wang}
\ead{hywang777@whut.edu.cn}

\author[label2,label3]{Adam J. Sobey}
\ead{ajs502@soton.ac.uk}

\cortext[cor1]{Corresponding author.}
\address[label1]{School of Transportation and Logistics Engineering, Wuhan University of Technology, YuJia Tou Campus No.1178 Heping Avenue Wuchang district, Wuhan, 430063, China}
\address[label2]{Maritime Engineering, University of Southampton, Southampton, SO17 1BJ, UK}
\address[label3]{Data-Centric Engineering, The Alan Turing Institute, The British Library, 96 Euston Road, London, NW1 2DB, UK}
\begin{abstract}
Supply networks require collaboration in a competitive environment. To achieve this, nodes in the network often form symbiotic relationships as they can be adversely effected by the closure of companies in the network, especially where products are niche. However,  balancing support for other nodes in the network against profit is challenging. Agents are increasingly being explored to define optimal strategies in these complex networks. However, to date much of the literature focuses on homogeneous agents where a single policy controls all of the nodes. This isn't realistic for many supply chains as this level of information sharing would require an exceptionally close relationship. This paper therefore compares the behaviour of this type of agent to a heterogeneous structure, where the agents each have separate polices, to solve the product ordering and pricing problem. An approach to reward sharing is developed that doesn't require sharing profit. The homogenous and heterogeneous agents exhibit different behaviours, with the homogenous retailer retaining high inventories and witnessing high levels of backlog while the heterogeneous agents show a typical order strategy. This leads to the heterogeneous agents mitigating the bullwhip effect whereas the homogenous agents do not. In the high demand environment, the agent architecture dominates performance with the Soft Actor-Critic (SAC) agents outperforming the Proximal Policy Optimisation (PPO) agents. Here, the factory controls the supply chain. In the low demand environment the homogenous agents outperform the heterogeneous agents. Control of the supply chain shifts significantly, with the retailer outperforming the factory by a significant margin.

\end{abstract}

\begin{keyword}
Multi-agent systems \sep Decision Support Systems \sep
Inventory Optimisation \sep Backlog and Stock-out  \sep Pricing
\end{keyword}

\end{frontmatter}

\section{
Agent based control of multi-echelon Supply Chain environments 
} 

Supply Chains (SC) are a network of suppliers, warehouses, distribution centres and retailers through which raw materials are acquired, transformed and delivered to customers. To model this complex scenario, the classical inventory control problem describes a decision-maker who must determine an order quantity in each period, such that the risk of over-ordering and under-ordering are balanced. The nodes must maintain the right balance between the supply and demand of products by optimizing ordering \citep{wang2023agent,leluc2023marlim,tian2024iacppo,wang2021spare,guo2023collaborative} and pricing \citep{yavuz2024deep,qiao2024distributed,alamdar2024deep} to strike a balance between stock availability and storage costs while minimising stockouts and overstocks \citep{toomey2000inventory,yang2023versatile}.

Each node in this supply chain is a self-interested agent, each of which are trying to stay financially viable, collaborating and competing with other nodes in the supply chain. The properties of different supply networks vary and the nodes need to determine which strategy might be most successful, from  monopolies which have high interdependencies with the echelon above and below where collaboration is vital, to networks with vast numbers of interconnected companies with large competition where the surrounding nodes might be easily replaceable and profitable companies might choose not to collaborate \citep{jiang2023quantile,dogan2015reinforcement}. To ensure profitability each partner in the supply chain increases or decreases pricing and inventory \citep{brintrup2010behaviour}. Over time each node evolves it's strategy to retain profitability, sometimes favouring more collaboration and sometimes competing against the companies in it's network. What makes the problem challenging is that in many cases a node will have limited or no information about the strategy that other companies in the network are following or what their inputs or outputs are, it's a Hidden Markov Decision Process. 

Increasingly, multi-agent modelling is being explored as a solution to control these supply chains. The literature is split into two main approaches to agent architecture, homogeneous and heterogeneous. Homogeneous approaches have been shown to performs well in various supply chain scenarios with a particular focus on inventory control, attempting to reduce the number of stockouts or backlogs \citep{stranieri2022deep,stranieri2024performance,kosasih2022reinforcement,wang2023agent,hubbs2020or,paine2022behaviorally,vanvuchelen2020use,gijsbrechts2022can, chen2021data, keskin2022data, shi2016nonparametric} . There is also a consideration of pricing and ordering decision making  
 \citep{yavuz2024deep,qiao2024distributed}, although the pricing isn't considered at the same time as stockouts or backlogs. Homogeneous approaches consider a single policy that supplies an action for each node in the supply chain and receives a reward. The homogeneous agents therefore have a shared observation, meaning that the observability of the environment is higher and the actions of the other actors can be optimised to work together \citep{arifouglu2010optimal}. However, for most supply chains, this level of sharing between companies would be unrealistic and for most real world scenarios, we will see agents developed separately by each node. This means that the observation space will be limited and the decision-making approach of other nodes will be hidden. In this case the heterogeneous literature seem a more realistic analogy to most multi-company supply chains.

\begin{table*}[!htb] %
	\centering %
\caption{State-of-the-art in multi-agent reinforcement learning for supply chain management; \textbf{On} indicates that on-Policy is considered; \textbf{Off} indicates that Off-Policy is considered; 
  \textbf{B} indicates that backlog is considered; \textbf{S} indicates that stockout is considered; \textbf{Partial} indicates that a partially-observable Markov Decision Process is considered; \textbf{Full} indicates that environment is fully observable and \textbf{Hidden} indicates that the decision making at the other nodes is Hidden.
} %
	\label{tab:Different approaches} %
	\fontsize{10}{9}\selectfont    %
	\begin{threeparttable} %
 \centering
  \setlength{\tabcolsep}{0.9mm}{
			\begin{tabular}[]{p{2cm}p{2cm}p{2cm}p{1.7cm}p{1.7cm}p{2.3cm}p{1cm}
   } 
				\toprule         
		\multicolumn{7}{c}{\bf Comparison of key setting in supply chain literature } \\ 
  \cmidrule{1-7}  
Reference  & Cooperation & Competition& Stockout, Backlog &Approaches& Observability  & Policy
 \\
	\midrule

\cite{kim2024multi}  & \checkmark  & &  &Hetero-Maximax Q-learning&Partial &Off\\
\cite{ding2022multi}    & &\checkmark &  S&CD-PPO&Partial/Hidden & On\\
\cite{sultana2020reinforcement}  & &&  S& A2C &Partial&On
 \\
 \\
\cite{yu2020multi}     & \checkmark  &  & &DDDQN &Full  &Off \\
\cite{liu2022multi}          & \checkmark   && B &HAPPO, PPO&Partial & On\\
\textbf{Ours}    & \checkmark &\checkmark  & B/S  &SAC, PPO &Partial/Hidden&  On/Off\\
 \bottomrule %
		\end{tabular}}
	\end{threeparttable}
\end{table*}

Table \ref{tab:Different approaches} summarises recent multi-agent reinforcement learning approaches focused on heterogeneous agents. There is more of a focus on cooperative supply chains \citep{kim2024multi, yu2020multi, liu2022multi} than competitive ones \citep{ding2022multi}. In \cite{kim2024multi} the focus is across different echelons of the supply chain, rather than between different echelons determining how competing companies might collaborate effectively. \cite{liu2022multi} and \cite{yu2020multi} explore how cooperation can be achieved across different echelons through profit sharing. \cite{liu2022multi} consider using an approach where the proportion of individual and group rewards can be adjusted, showing better performance when the agents are more selfish.  However, we see limited use of profit sharing in real supply chains making it unrealistic for most practical applications, and alternative reward sharing approaches need to be developed. In competitive environments, \cite{ding2022multi} investigate a single store with multiple stock keeping units, with a shared inventory. In this case each of the stock keeping units has an individual reward to provide but overstocking is reduced by proportional reduction of the excess inventory back to the capacity. This multi-agent literature  focuses on either scenarios with stock-outs \citep{ding2022multi,sultana2020reinforcement} or avoiding backlog \citep{mousa2024analysis,liu2022multi,yang2023versatile} with no literature considering both together. However, this allows easy solutions. For example, if an agent needs to avoid stockout but there is no backlog, then the agent can keep the inventory at the maximum value with no penalty.

Therefore, this paper compares homogeneous and heterogeneous approaches to supply chain management to determine whether the observability and transparency of decision making changes the feasibility of multi-agent controlled supply chains resulting in a more co-operative partnership. It does this in high demand and low demand scenarios where agents are required to make decisions on order quantities and pricing, as well as simultaneously avoiding backlogs and stockouts. An approach to reward sharing is developed, where the agents are penalized for the other agents running out of stock but where there is no profit sharing. On-policy and off-policy algorithms are compared, multi-agent PPO as the on-policy algorithm and multi-agent SAC as the off-policy.

\section{Heterogeneous Hidden Markov Supply Chain Environment}
In a multi-agent system, each agent has its own observations, actions, and rewards. We can denote a multi-agent reinforcement learner with the tuple $(N,S_N,Obs_N,A_N,P_N,R_N)$ in which $N$ is the total number of the agents; $S_N ={s_{1},\dots,s_{i}}$ is the state space for each agent;  $Obs_N={obs_{1},\dots,obs_{i}}$ is the set of observations for each agent; $A_N ={a_{1},\dots,a_{i}} $ is the action space for each agent; $P_N(s_i'|s_i,a_i)$ denotes the transition probability from $s$ to $s'$ from all $i$ agents and $R_{N}= r_1(obs_{1,t},a_{1,t},obs_{1,t+1})...r_i(obs_{i,t},a_{i,t},obs_{i,t+1})$ denotes agent $i$ takes action $a_{i,t}$ given observation $obs_{i,t}$ at time-step $t$ and then receives an immediate reward $r_{i,t}$ and a new observation $obs_{i,t+1}$.

\begin{figure*}[h]
	\centering
	\includegraphics[width=0.65\linewidth]{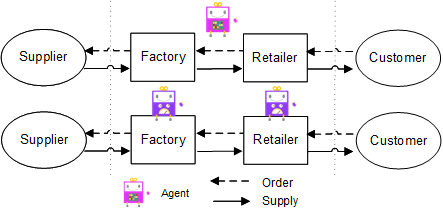}
 \caption{Multi-agent approach to solving an inventory dynamics model in a two-echelon supply chain.}
	\label{fig:sc structure}
\end{figure*}

A two echelon supply chain is constructed, to reduce the complexity in the environment and to help understand the behaviour at each node. The multi-echelon multi-agent supply chain model is shown in Fig.\ref{fig:sc structure} made up of one factory agent and one retailer agent. The upstream factory agent supplies intermediate products to the retailer agent which provides final products to satisfy customer demand. The factory is assumed to be able to buy as much stock as ordered and the retailer must meet the demand specified by the customer. Two demand scenarios are tested one with a high demand $(D \sim Poisson(\mu = 10)$ and one with a low demand $ D \sim Normal(mean=2,std=1))$. 

Two agent architectures are compared on these scenarios, a homogeneous agent which represents the approach taken across most of the literature where there is a single agent with the same observability for all agents and a heterogeneous agent where different elements of the supply chain are represented by agents  which have separate observations. Table \ref{Tab:Variables} summarises the environment parameters for the multi-echelon supply chain configurations used and the source code is released at  GitHub \footnote{\url{https://github.com/wangwan0910/masc}}.

\begin{table*}[!htb]\centering					
	\caption{Agent parameters in two-echelon supply chain.}				
	\label{Tab:Variables}				
	\begin{tabular}[]{p{1.2cm}p{4.4cm}p{1.5cm}p{1.5cm}}\toprule				
		\textit{Notation} & \textit{Explanation}& Retailer (i=1)&  Factory (i=2) \\ \midrule			
$Sp_{i}$  & Unit Sales Price &[0,6] &[0,6] \\	
$Q_{i}$& Purchasing Quantity(Real Order) & [0,20]&[0,20] \\	
$c_{i}$  & Unit Purchasing/Ordering Cost & \textbf{$Sp_{2}$}  &   0.2 \\
$Hc_{i}$ & Unit Inventory Holding Cost &0.2&0.2  \\	
 $I_{i}$ & Initial Inventory Level &10&10 \\	
$C_{i}$ & Inventory Capacity  &19&59  \\
\textbf{$Sc_{i}$} & Unit Stockout Cost &140&70  \\	
$SL_{i}$& Initial Stockout Level & 0&0 \\
 $Bc_{i}$ & Unit Backlog Cost &  1&    1  \\
 $B_{i}$ & Initial Backlog Level  &  0&    0  \\
$D $& Customer Demand &D&$Q_{1}$   \\	

$T$  & Simulation Months &  30 &  30\\
\bottomrule			
\end{tabular}				
\end{table*}	

\subsection{Mathematical Formulation}
\label{Mathematical Formulation}

The two agents: retailer $i=1$, and factory $i=2$, maximise their rewards by minimising the sum of the inventory and stock-out costs, shown in  Eq. \ref{E2},  

\begin{equation}
\label{E2}
\begin{aligned}
\max & \sum_{t=0}^{T}  \begin{bmatrix} 
\underbrace{Sp_{i} \times \sum_{0}^{i}Q_{i}}_{\text {Total nodal profit}}-\underbrace{  Hc_{i} \times I_{i}}_{\text {Total nodal inventory cost}}
\\- \underbrace{Bc_{i} \times  max(I_{i} - C_{i},0)}_{\text {Total nodal backlog cost}}-
\underbrace{  Sc_{i} \times  max(Q_{i,t}-I_{i},0)}_{\text {Total nodal stockout cost}}
\\-\underbrace{ c_{i+1} \times Q_{i}}_{\text {Total nodal purchasing cost}}
\\
\end{bmatrix}\\ \text{subject to:} \\ 
& i \in {1,2},\\
& 0 \le Q_{i,t} \le C_{i}, \\
\end{aligned}
\end{equation}

where $Sp_{i}$ is the unit sales price for each agent, $i$; $Hc_{i}$ signifies the inventory holding cost; $I_{i,t}$, represents the inventory level at each echelon of the supply chain at a specific time $t$; 
$D_{t}$ represents the demand at the customer and $C_{i}$ reflects the maximum inventory capacity for each agent. This is subject to the inventory capacity constraints which are 20 for the retailer, and 60 for the factory. 

$Sc_{i}$ represents the stock-out cost when the node is out of stock and the total nodal stockout cost: $ TS_{c} = -Sc_{i} \times  max(Q_{i,t}-I_{i},0)$;where $Bc_{i}$ represents the backlog cost, when the node is over the maximum stock and the total nodal backlog cost: $ TB_{c} = Bc_{i} \times  max(I_{i} - C_{i},0)$.

\subsubsection{State space}
The state $s_{i,t}$ is defined as a vector in  Eq.\ref{Es},

\begin{equation}
\label{Es}
s_{i,t}=\left \{ s_{1,t} ,s_{2,t} 
\right \} , i\in \left\{ 1,2  \right\}, t\in \left \{ 1,\cdots T  \right \},
\end{equation}

where the state for the retailer is defined in Eq. \ref{State0},

\begin{equation}
\label{State0}
s_{1,t}=\left \{I_{1,t},B_{1,t},SL_{1,t},D_{1,t-2},D_{1,t-1},D_{1,t},p_{t}| i\in \left\{ 1,2  \right\}, t\in \left \{ 1,\cdots T  \right \}
\right \} ,
\end{equation}

and the  state for the factory is defined in Eq. \ref{State1},

\begin{equation}
\label{State1}
s_{2,t}=\left \{I_{2,t},B_{2,t}, SL_{2,t},D_{2,t-2},D_{2,t-1},D_{2,t},p_{t}| i\in \left\{ 1,2  \right\}, t\in \left \{ 1,\cdots T  \right \}
\right \}. 
\end{equation}

\subsubsection{Action space}

In the simple Markov model, all states are observable. However, in many real world scenarios some of the observations are not available to the agent, referred to as Partial Observable Markov Decision Processes (POMDP). In addition to this the observations seen by an agent might be determined by a Markov Process, hidden from the agent. In this environment the homogeneous agents gain the same observation as they share a single policy, shown in Eq. \ref{Ea1}, 

\begin{equation}
\label{Ea1}
\begin{aligned}
obs_{i,t} = \{I_{1,t},I_{2,t},B_{1,t},B_{2,t},SL_{1,t},SL_{2,t},D_{1,t-2},D_{2,t-2},\\
D_{1,t-1},D_{2,t-1},D_{1,t},D_{2,t},p_{t}
 \mid i\in \{ 1,2  \}, t\in  \{ 1,\cdots T   \} \}. 
 \end{aligned}
\end{equation}

However, the heterogeneous agents witness different observations, the retailer has a partial view of the world with no ability to see what the factory can observe. The retailer's decision making process is hidden from the factory.  In each period $t$, agent $i$ observes the new and previous demand, defined in Eq. \ref{Ea2}, 

\begin{equation}
\label{Ea2}
obs_{i,t} = \{I_{i,t},B_{i,t},SL_{i,t},D_{i,t-2},D_{i,t-1},D_{i,t},
p_{t}|i\in \left\{ 1,2  \right\}, t\in \left \{ 1,\cdots T  \right \} \}. 
\end{equation}

In this environment the homogeneous agents gain the same action as they share a single action space, shown in Eq. \ref{Eah}, 

\begin{equation}
\label{Eah}
a_{i,t}=\{  Q_{1,t} , Q_{2,t},Sp_{1,t}, Sp_{2,t}
  | i\in \left\{ 1,2  \right\}, t\in \left \{ 1,\cdots T  \right \}\}.
\end{equation}

For the heteorgeneous agents, the action $a_{i,t}$ is defined in  Eq. \ref{Ea} as a vector of ordering and  pricing,

\begin{equation}
\label{Ea}
a_{i,t}=\left \{ a_{1,t} ,a_{2,t} 
\right \} , i\in \left\{ 1,2  \right\}, t\in \left \{ 1,\cdots T  \right \},
\end{equation}

with the retailer being able to order product, $Q_{1,t}$ and set the price for the product,
defined in Eq. \ref{E9},

\begin{equation}
\label{E9}
a_{1,t}=\left \{ Q_{1,t} , Sp_{1,t}
\right \},
\end{equation}

and the factory having the same actions, defined in Eq. \ref{E10},

\begin{equation}
\label{E10}
a_{2,t}=\left \{ Q_{2,t}, Sp_{2,t}\right \}.
\end{equation}

\subsubsection{State transition}
The transition function is implemented according to the material balance constraints in Eq. \ref{E6:transition},

\begin{gather}
\label{E6:transition}
\begin{aligned}
I_{i,t+1} &= I_{i,t} + Q_{i,t} - D_{i,t}, i\in \left\{ 1,2  \right\}.\\
\end{aligned}
\end{gather}

The downstream participants' demand is the upstream participants' order. The inventory at the next step is the addition of the current orders to the previous step and the subtraction of current demand. 

\subsection{Reward Function}
The goal of the decision-maker in the inventory control problem is to balance the risk of over-ordering and under-ordering by determining the optimal pricing and order quantity in each ordering period. If the agent chooses a certain action through trial and error at time step t, then the reward, $r_{i,t}$ can be calculated using Eq. \ref{Er:r1},

\begin{equation}
\label{Er:r1}
r_{i,t+1} =\left \{ r_{1,t+1},r_{2,t+1}\right \},
\end{equation}

where the reward for the retailer is defined in Eq. \ref{E:reward1},
\begin{equation}
\label{E:reward1}
\begin{aligned}
\text{$r_{1,t}$} = 
& Sp_{1} \times D - 0.2 \times I_{1,t}
 - \max(I_{1,t} - 20, 0) -   140 \times \max{(D - I_{1,t}, 0)} -  Sp_{2} \times Q_{1,t},
\end{aligned}
\end{equation}

and where the reward for the factory is given in Eq. \ref{E:reward2}, 
\begin{equation}
\label{E:reward2}
\begin{aligned}
\text{$r_{2,t}$} = &  Sp_{2} \times Q_{1}  -  0.2 \times I_{2,t} -  \max(I_{2,t} - 60, 0) 
-  70 \times \max{(Q_{1,t} - I_{2,t}, 0)}  -  0.2\times Q_{2,t} .
\end{aligned}
\end{equation}

In this Baseline environment the agents will learn to choose the optimal action at each state to maximize the agent's profits $r_{i,t}$. However, in many supply chains it will be important to collaborate to ensure that niche suppliers or customers do not go out of business. In the previous literature this has been done by profit sharing \cite{oroojlooyjadid2022deep} and \cite{mousa2024analysis}, but it seems unlikely that this mechanism would be realistic for all but the most integrated supply chains. Therefore, reward shaping is introduced where the agents are penalised for the other agent running out of stock; this is defined in  Eq. \ref{CollaR},

\begin{equation}
\label{CollaR}
\begin{aligned}
r_{i,t+1} =\left \{ r_{1,t+1} -70 \times \max{(\textbf{Q}_{1,t} - I_{2,t}, 0) },r_{2,t+1}  -140 \times \max{(\textbf{D} - \textbf{I}_{1,t} , 0)  }\right \}.
\end{aligned}
\end{equation}





\subsection{ Multi-agent reinforcement learning experimental settings}
\label{experimental settings}

The environment is based on the OpenAI Gym APIs framework \cite{brockman2016openai} and Ray's multiagent tools for simulating multi-echelon, multi-agent supply chain environments.  Each agent's neural network was built, compiled and trained using Pytorch. All experiments were run on the IRIDIS supercomputer (SLURM, 2023) using CPU Cores Intel(R) Xeon(R) E5-2670, and GPUs (NVIDIA Quadro RTX8000). Hyperparameters play a crucial role in the context of multi-agent deep reinforcement learning algorithms since they can significantly influence training and, consequently, relative performance. The multi-agent algorithms are turned using Ray Tune \citep{moritz2018ray}, a scalable hyperparameter tuning library, which is an open-source library Rllib \citep{liang2018rllib}. Appendix Table \ref{Tab:PPO Hyperpara}, \ref{Tab:SAC Hyperpara} lists the  selected hyperparameters for the homogeneous and heterogeneous agents.

\section{Comparison of homogeneous and heterogeneous agents in the high demand environment}
A numerical analysis is performed to compare the performance of homogeneous and heterogeneous agents using SAC and PPO agents. This is followed by an exploration of whether cooperation can be generated through reward sharing without profit sharing and whether this is detrimental to the actors.

\subsection{Comparisons of the homogeneous and heterogeneous agents on the high demand baseline environment} 
\begin{figure*}[!htbp]
    \centering
    \begin{subfigure}[!htbp]{0.485\linewidth}
        \centering
\includegraphics[width=\linewidth]{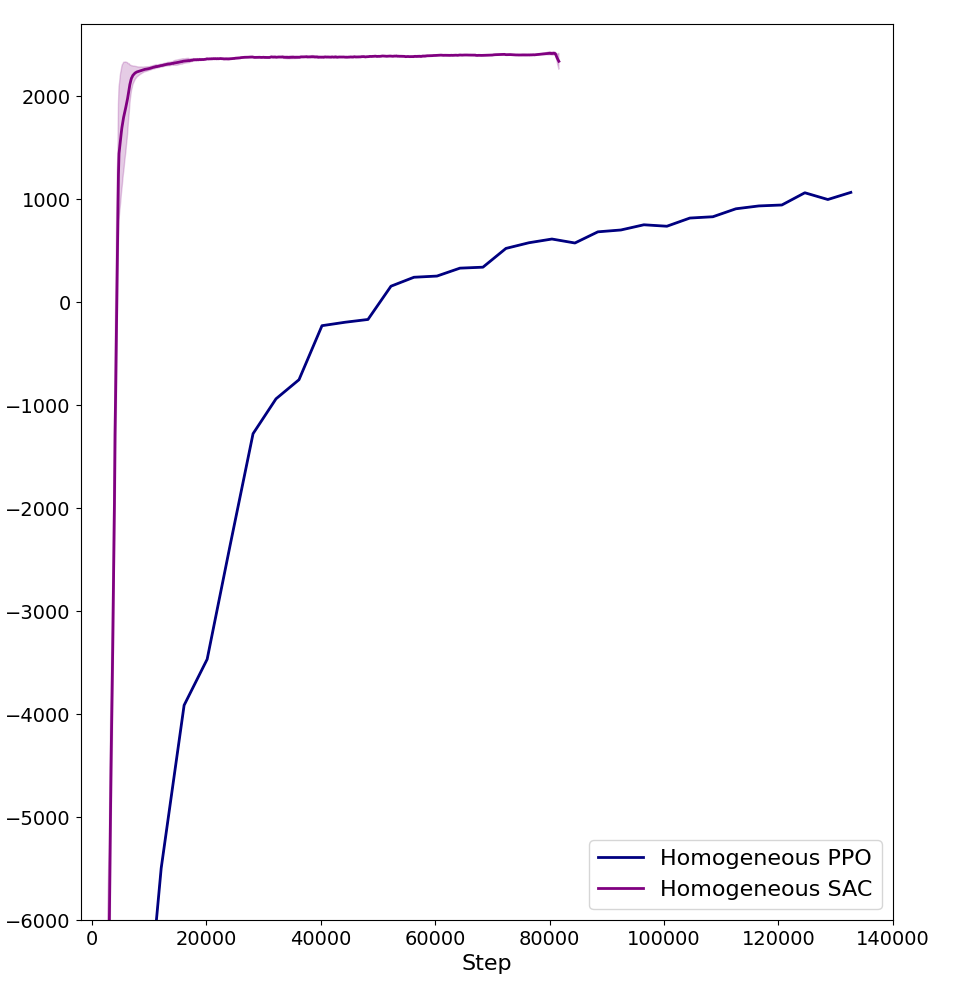}
  \subcaption{Homogeneous agent.}
    \label{fig:episode}
  \end{subfigure}
    \begin{subfigure}[!htbp]{0.485\linewidth}
        \centering
\includegraphics[width=\linewidth]{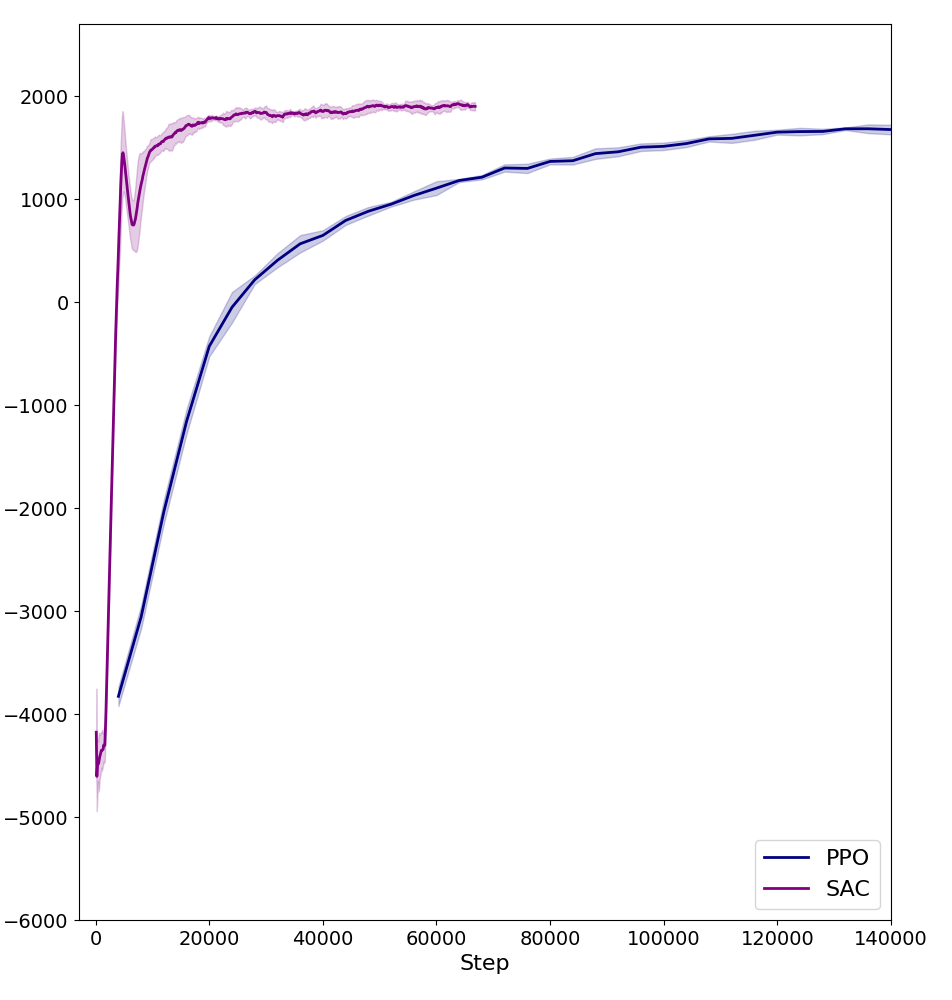}
  \subcaption{Heterogeneous agent.}
    \label{fig:episode}
  \end{subfigure}

\caption{Comparison of agent's performance in homogeneous and heterogeneous configurations, using PPO and SAC architectures in the high demand environment. The shaded area depicts the standard deviation of the multi-agent's performance for independent experiments using 5 different seeds.}
\label{fig:episode-reward-high}
\end{figure*}

Fig. \ref{fig:episode-reward-high} shows that the SAC rewards are higher than PPO for both homogeneous and heterogeneous agents. The most profitable is the homogeneous SAC agent, which has a profit of 2,406 per episode, while the homogeneous PPO agent generates 1,271 an episode. For the heterogeneous agent the SAC's reward reaches 1,891, lower than the homogeneous SAC agents reward, while the PPO's reward is 1,754, higher than the homogeneous PPO configuration. There is a limited variation between the 5 simulations for any of the 4 different architectures, with the highest variation early in the SAC training.

\begin{figure*}[!htbp]
    \centering
    \begin{subfigure}[!htbp]{0.46\linewidth}
        \centering
\includegraphics[width=\linewidth]{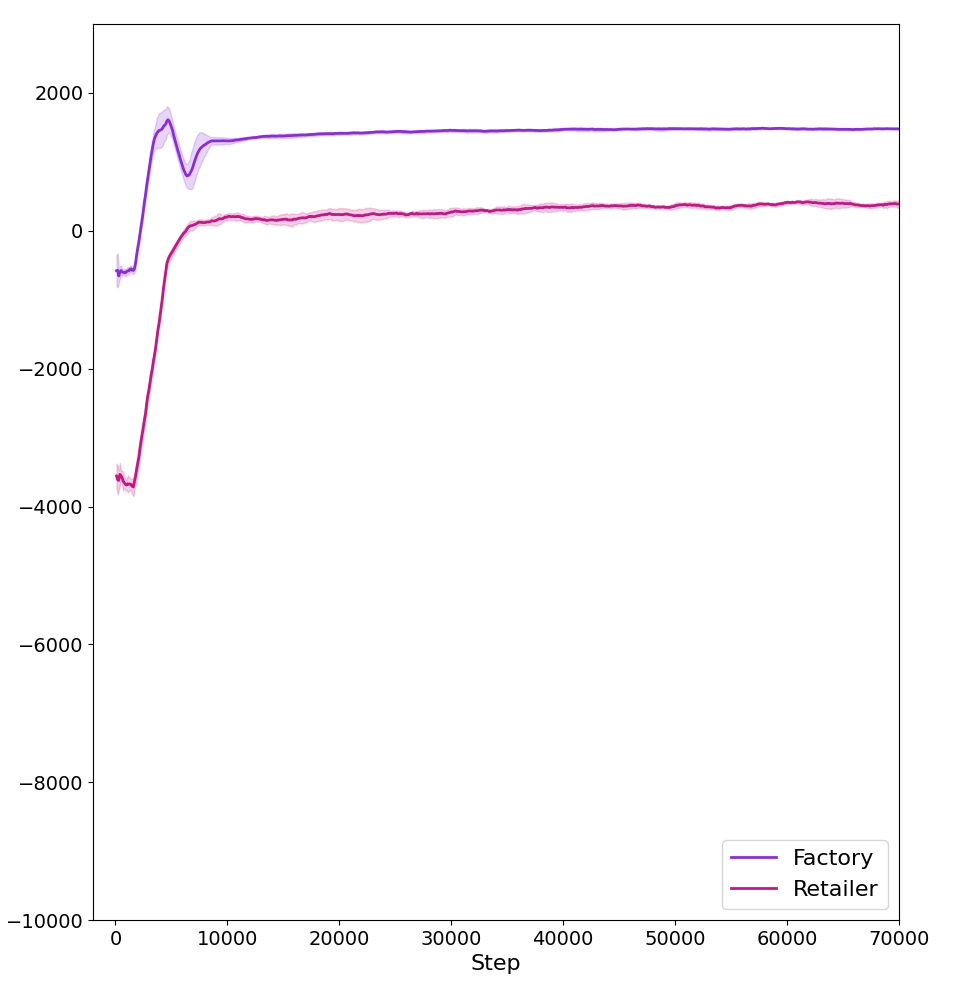}
  \subcaption{SAC}
    \label{fig:episode}
  \end{subfigure}
    \begin{subfigure}[!htbp]{0.46\linewidth}
        \centering
\includegraphics[width=\linewidth]{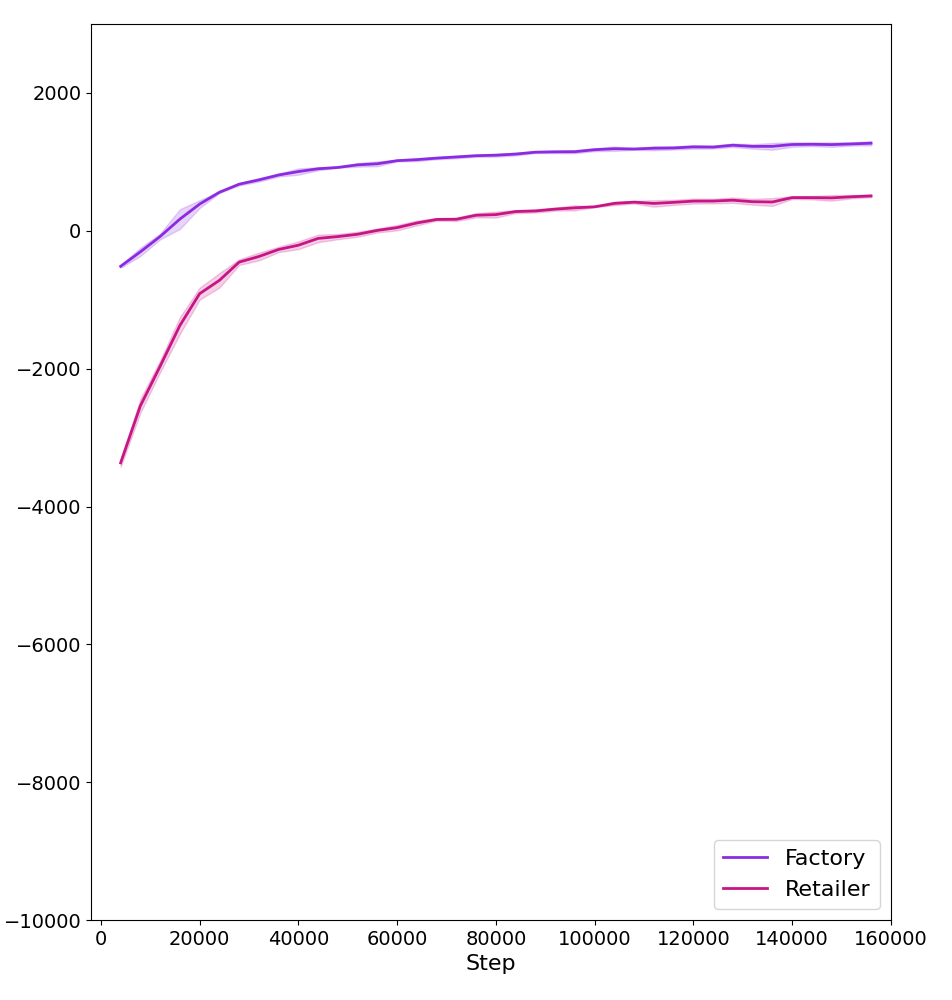}
  \subcaption{PPO}
    \label{fig:episode}
  \end{subfigure}
\caption{Comparison of the factory and the retailer agent's reward for the SAC and PPO architectures in the high demand environment. The shaded area depicts the standard deviation of the heterogeneous agent's performance for independent experiments using 5 different seeds.}
\label{fig:factory-retailer-reward-high1}
\end{figure*}

When comparing the behaviour of the factory and the retailer for the heterogeneous agents, Fig. \ref{fig:factory-retailer-reward-high1} shows that for both the PPO or SAC algorithms the factory profit is always higher than the retailer profit.  After training the SAC factory agent has a reward of 1,497 per episode while the retailer has a reward of 394 per episode. For the PPO agent then the factory has a reward of  1,260 per episode while the retailer has a reward of 494 per episode. In all of the scenarios, the variation between the 5 repeats is again limited, reflecting the overall performance. 

\begin{figure*}[!htbp]
        \centering
        \includegraphics[width=\linewidth, height=6.4cm]{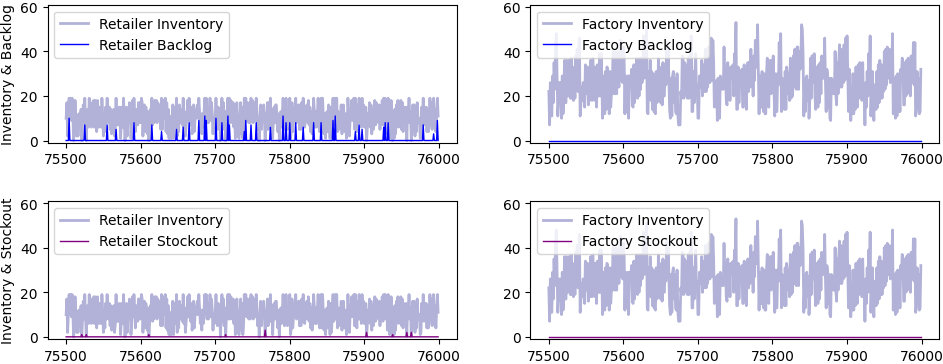} 
     \caption{Heterogeneous SAC agent actions and resulting backlog and stockouts in the high demand scenario. The mean inventory of the retailer, on the left, is 11.1 while the mean inventory of the factory, on the right, is 28.}
     \label{fig:stockout-backlog high1}
\end{figure*}

Comparing the strategy for the heterogeneous and homogeneous agents, the SAC agents are compared as these agents have a higher performance. Fig.\ref{fig:stockout-backlog high1} shows a representative example of 500 days of inventory, stockout count and backlog count. The heterogeneous retailer agent has a highly fluctuating stock level that adjusts to be in the middle of the capacity but that stretches from maximum capacity to empty with a mean inventory of 11.1. In this case there are irregular and limited numbers of stockouts but a large number of backlogs that occur regularly. The factory in this environment, has a more regular pattern buying stock and then letting the inventory reduce but the overall inventory level remains high with a mean of 28. The inventory never reaches the maximum value or the minimum value and there are no stockouts or backlogs over this period. This aligns well with the procedure established by \cite{hekimouglu2018markov} to save costs and mitigate supply risks with regular orders.

\begin{figure*}[!htbp]
    \centering
        \includegraphics[width=\linewidth, height=6.4cm]{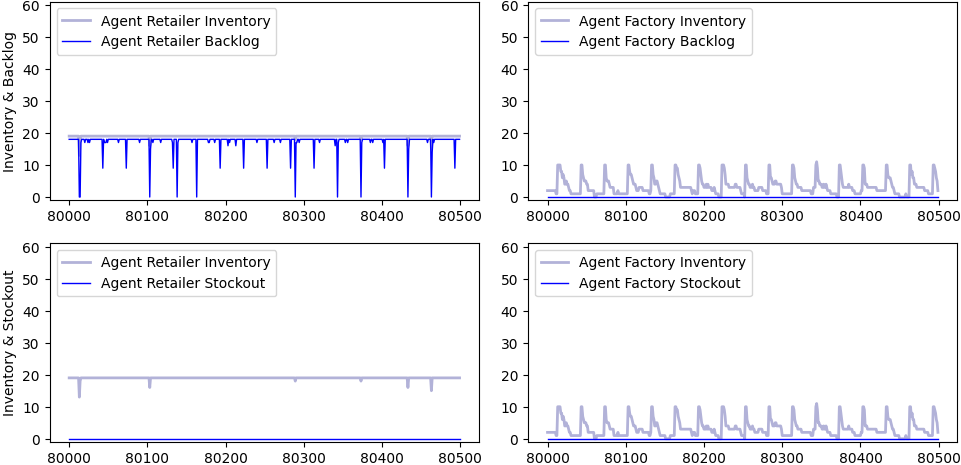}
    \caption{Homogeneous SAC agent actions and resulting backlog and stockout in the high demand scenario. The inventory of the factory is 3.13 while the inventory of the retailer is 18.91.}
    \label{fig:stockout-backlog-high2}
\end{figure*}

The homogeneous agent, Fig. \ref{fig:stockout-backlog-high2} uses a different strategy, with a high inventory level at the retailer, constantly near the maximum of 19 at 18.91, this invokes a high inventory and backlog cost and the agent takes on the maximum backlog penalty almost every day. However, these costs are low  with inventory costing 0.2 per unit and backlog costs of 1 per unit while the stockout cost is 140 and by keeping the stock high stockouts are avoided. However, the factory follows a different strategy, making orders at regular intervals and letting the stock drop gradually over time. This more cloesly follows the standard inventory buying strategy. The peaks in the inventory level are much lower than in the heterogeneous agent factory, reaching a maximum of about 10. In this case the stockout cost is 70 and so there is less of a risk in lower stock levels but the agent does not suffer a backlog or stockout penalty during the 500 days and mitigates the bullwhip effect, which is harder to do in the heterogeneous hidden Markov Decision Process.

\begin{figure*}[!htbp]
        \centering
        \includegraphics[width=0.75\linewidth, height=6.3cm]{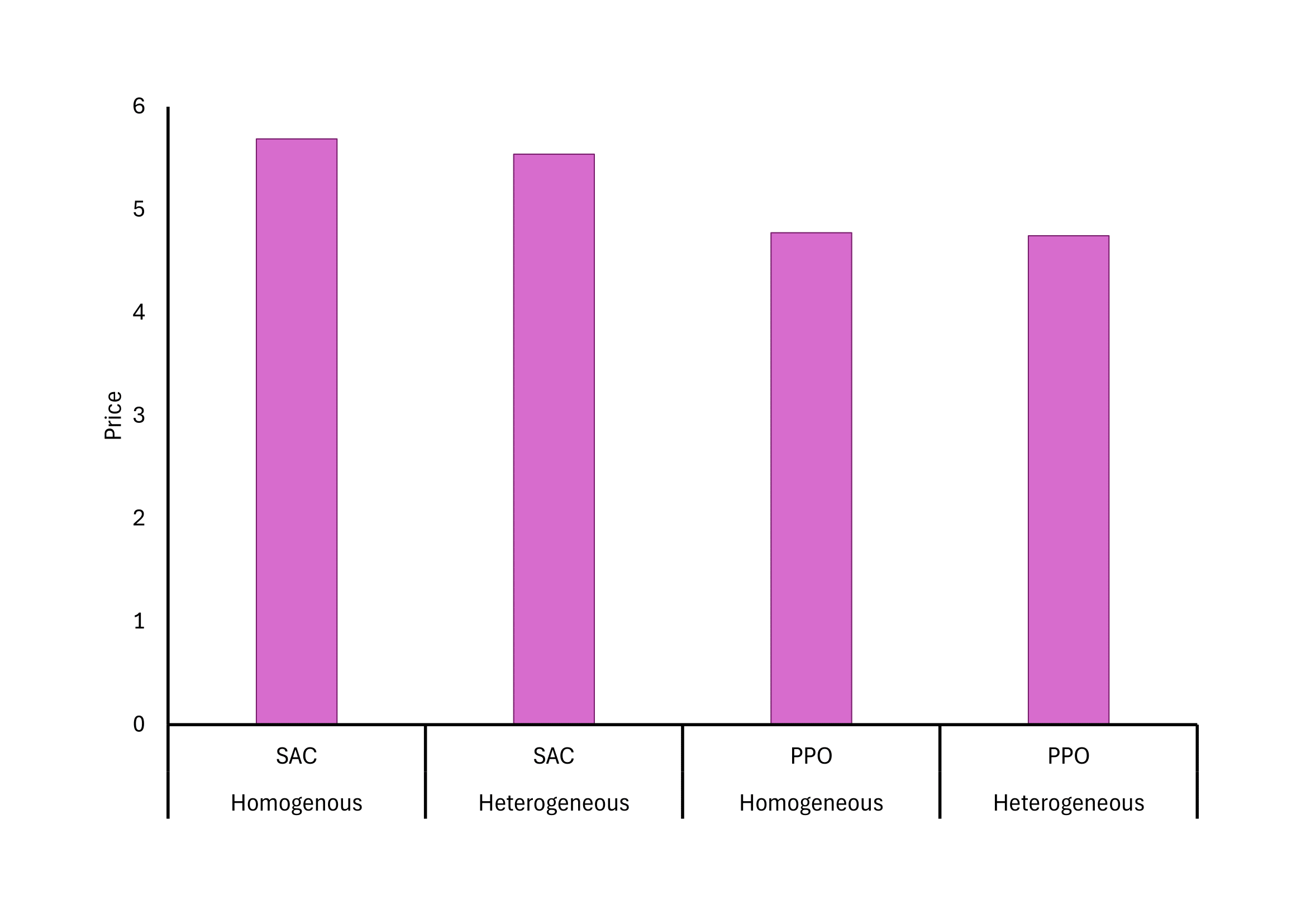} 
     \caption{Comparison of factory selling price of different architectural agents  in the high demand environment.}
     \label{fig:High-Demand-Price}
\end{figure*}

The price for the different agents shows a clear relation to the performance, with a comparison of the mean price in Fig. \ref{fig:High-Demand-Price}. The SAC agents charge a higher price than the PPO agents in both the homogeneous and the heterogeneous formats. The SAC agents charge almost the maximum of 6, with the homogeneous agent setting the mean price at 5.69 and for the heterogeneous agent it is 5.54. The heterogeneous agent shows a relatively consistent selection of these high values, but still occasionally selects values below 4. The PPO agent sets prices at a substantially lower mean value of 4.55 for the heterogeneous agent and 4.78 for the homogeneous agent. Both of these agents follow a similar trend in the price, selecting a value of 4 to 6 most rounds but there is more variation and the agents often select values lower than 4.

\subsection{Reward shaping to increase collaboration between agents in the high demand environment}
\label{the performance1} 

The reward shaping environments have a similar performance to the previous learning curves, except there is a larger variation in reward for the SAC agent in the heterogeneous configuration. In these cases the homogeneous SAC performs the best and the homogeneous PPO performs the worst. For the heterogeneous agent the mean episode reward value is 1,821 for the SAC algorithm in Collaboration, compared to 1,599 for the PPO algorithm in Collaboration. For the homogeneous agent, the mean episode reward value is 2,405 for SAC algorithm in Collaboration, compared to 1,297 for the PPO algorithm in Collaboration. The SAC values are similar to the baseline environment, despite the additional penalty applied to the reward. The heterogeneous PPO agent is substantially lower than the baseline, 1,599 compared to 1,754 but the homogeneous PPO agent performing slightly better, 1,297 compared to 1,271. However, it is more challenging to match the reward with the reward shaping, as a penalty is given for the performance of the other agent and so a double penalty is given on the total reward. Indicating a small improvement in performance.

\begin{figure*}[!htbp]
    \centering
          \begin{subfigure}[!htbp]{0.48\linewidth}
        \centering
        \includegraphics[width=\linewidth]{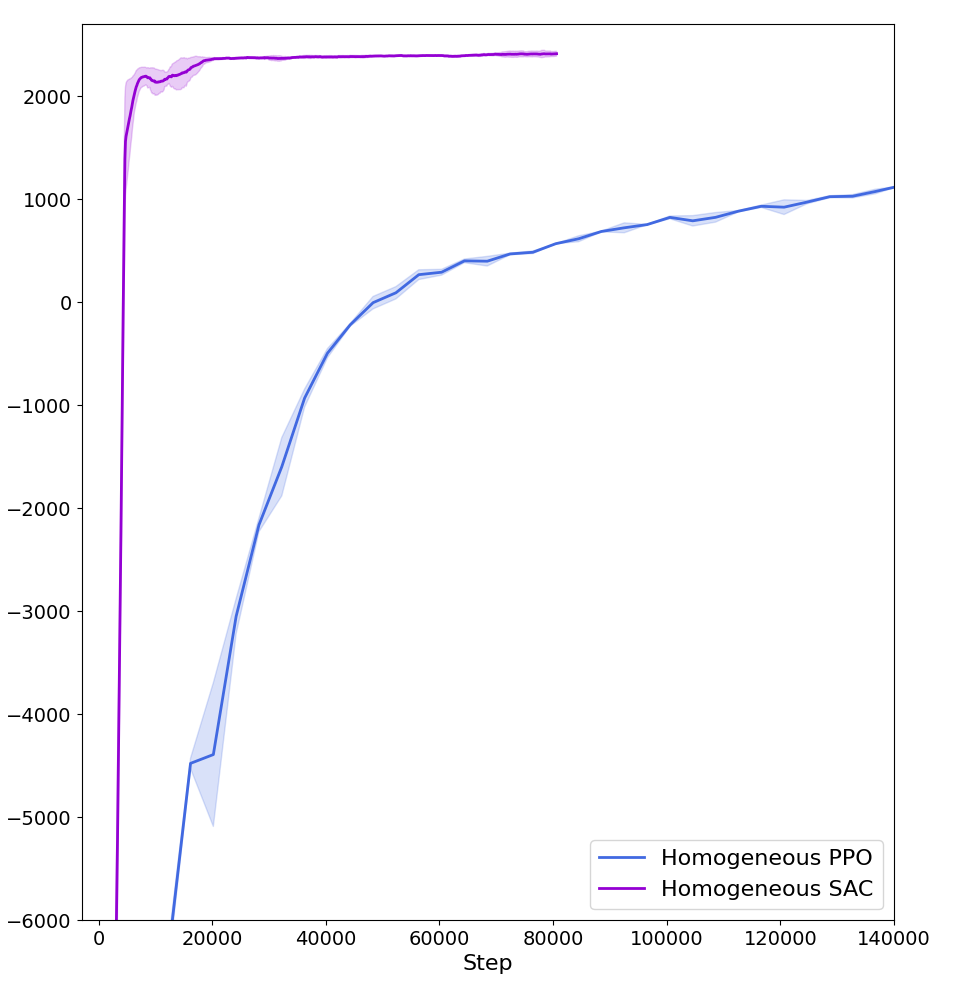}
        \subcaption{Homogeneous agent with Collaborative reward}
        \label{fig:policy4}
    \end{subfigure}
        \begin{subfigure}[!htbp]{0.48\linewidth}
        \centering
        \includegraphics[width=\linewidth]{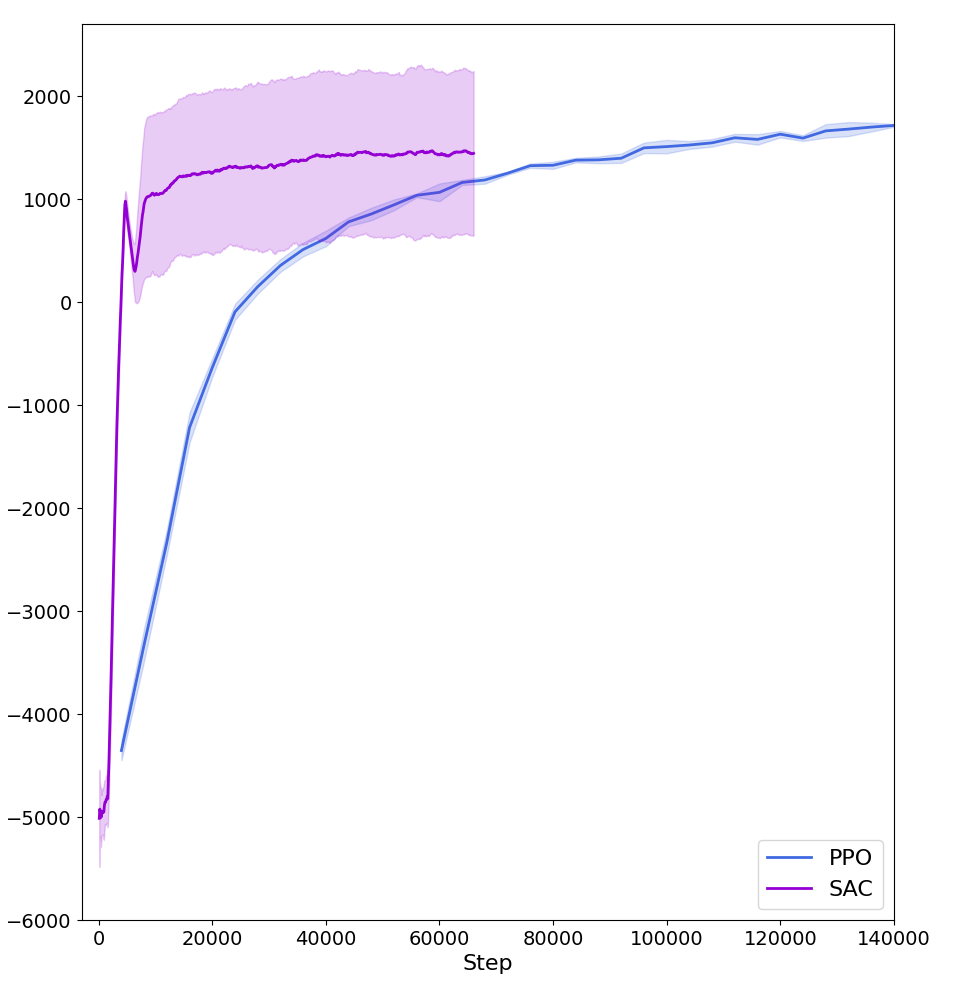}
        \subcaption{Heterogeneous agent with Collaborative reward}
        \label{fig:policy4}
    \end{subfigure}

    \caption{Comparison of the agent's performance with reward shaping in the homogeneous and heterogeneous configurations, using PPO and SAC architectures in the high demand environment. The shaded area depicts the standard deviation of the multi-agent's performance for independent experiments using 5 different seeds. }
    \label{fig:reward_high_23}
\end{figure*}

\begin{figure*}[!htbp]
    \centering
     \begin{subfigure}[!htbp]{0.46\linewidth}
          \centering
\includegraphics[width=\linewidth]{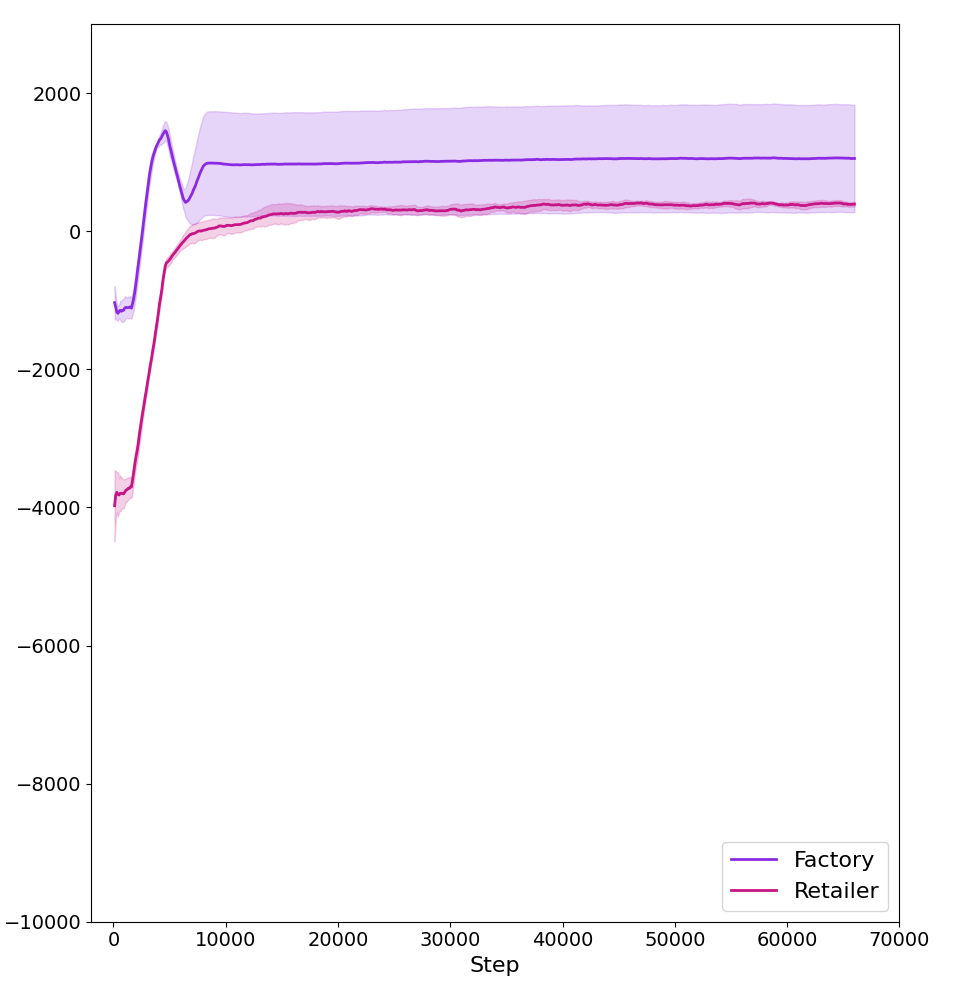}
  \subcaption{SAC with Collaborative reward shaping}
    \label{fig:episode}
  \end{subfigure}
    \begin{subfigure}[!htbp]{0.46\linewidth}
        \centering
\includegraphics[width=\linewidth]{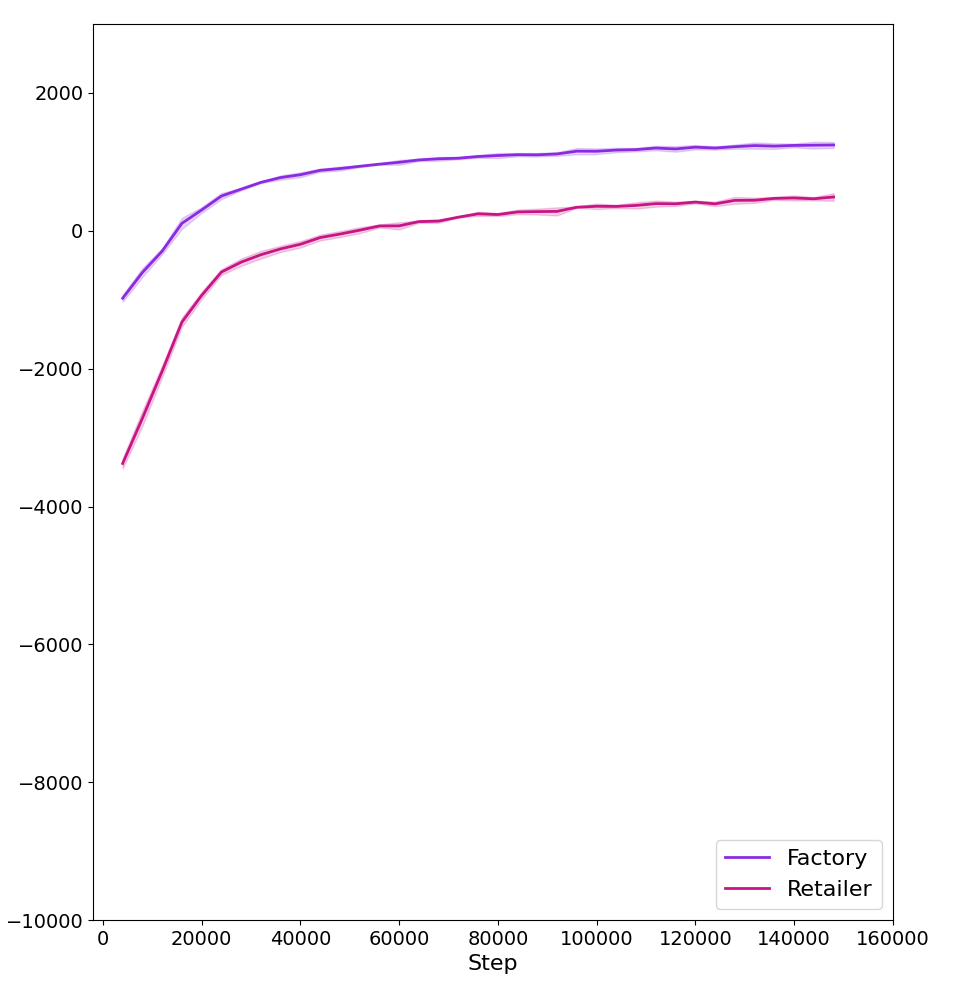}
  \subcaption{PPO with Collaboartive reward shaping}
    \label{fig:episode}
  \end{subfigure}
\caption{Comparison of the factory and the retailer agent's reward for the SAC and PPO architectures in the high demand environment. The shaded area depicts the standard deviation of the heterogeneous agent's performance for independent experiments using 5 different seeds. }
\label{fig:factory-retailer-reward-high23}
\end{figure*}

 Fig. \ref{fig:factory-retailer-reward-high23} provides the learning curves for the factory agent and retailer agents in the Collaborative environment. The curves follow a similar pattern to the baseline results except that the factory shows a larger variation in behaviour, explaining the total reward variation. For the SAC agent the Collaborative retailer's profit is 376 while the Collaborative factory agent's profit is 1,446. PPO follows the same behaviour with the Collaborative retailer agent  getting a reward of 423 and the Collaborative factory reward is 1,176. The SAC agent shows a wider separation of retailer and factory profit than the PPO agent.

For the inventory, then the strategy remains the same as for the baseline environment. However, there are some small changes in the mean inventory, shown in Table \ref{Tab:collainv}. Here the retailer generally has a lower inventory in the collaborative environment and the factory is lower in the baseline. This shows a shift in strategy to retaining more stock at the factory. However, in the PPO agent, which has a lower performance, the heterogeneous agent has higher mean stock in the collaborative environment for both nodes but the homogeneous agent has the lower mean stock at both nodes in the collaborative reward system. 

 \begin{table}[!htb]
     \centering
   \caption{Comparison of inventory levels of two symbiotic agents, Benchmark and Collaborative scenarios, in a high demand supply chain.}
     \label{Tab:collainv}
    \begin{tabular}{p{2cm}p{1.5cm}p{2cm}p{2cm}p{2cm}p{2cm}}
         \toprule
      Architecture & Agent & Baseline Factory & Baseline Retailer& Collaboration Factory & Collaboration Retailer\\
        \midrule
Homogenous	&SAC	&\bf{18.908}&	3.126&18.964&	\bf{2.564}\\
Heterogeneous	&SAC&		\bf{30.71}	&10.41&31.3	&\bf{10.12}\\
Heterogeneous	&PPO	&\bf{28.1}&	\bf{11.14}&29.99&	11.25\\
Homogeneous &	PPO	&	28.63	&15.81&\bf{24.33}&	\bf{14.44}\\
         \bottomrule
    \end{tabular}
 \end{table}
 
The prices in the collaborative environments show limited differences to those in the standard environment, the homogeneous SAC agent is 5.7 compared to 5.69 in the baseline, while the heterogeneous SAC agent has a mean of 5.66 compared to 5.54 in the baseline. For the PPO agent then the heterogeneous agent has a mean of 4.53 compared to 4.75 in the baseline and the homogeneous PPO agent has 4.68 compared to 4.78 in the baseline.

The high demand environment leads to different behaviours between agents. In both the baseline and in the collaborative reward sharing the SAC agents outperform the PPO agents. The SAC homogeneous agent performs better than the SAC heterogeneous agent but this is reversed for PPO. The homogeneous SAC retailer strategy is to retain a high inventory which leads to a large quantity of backlog incidents and a factory inventory that can be kept at a low level. However, the heterogeneous agent uses a strategy that has a reasonable number of backlogs at the retailer but where the factory can vary the stock to avoid backlog and stockout. The reward sharing, shows a shift in inventory from the retailer to the factory in the SAC agents and retains a similar reward despite this environment being more challenging. 

\section{Comparison of homogeneous and heterogeneous agents in the low demand environment}

A numerical analysis is conducted to compare the performance of Homogeneous and Heterogeneous agents on an environment with a lower demand to see how the agent architectures affects the buy strategies. First, the behavior of the SAC and PPO agents is assessed on the Baseline environment, followed by efforts to improve cooperation through reward sharing without sharing profit.

\subsection{Comparisons of the homogeneous and heterogeneous agents on the low demand baseline environment}
\begin{figure*}[!htbp]
    \centering
 \begin{subfigure}[!htbp]{0.485\linewidth}
        \centering
\includegraphics[width=\linewidth]{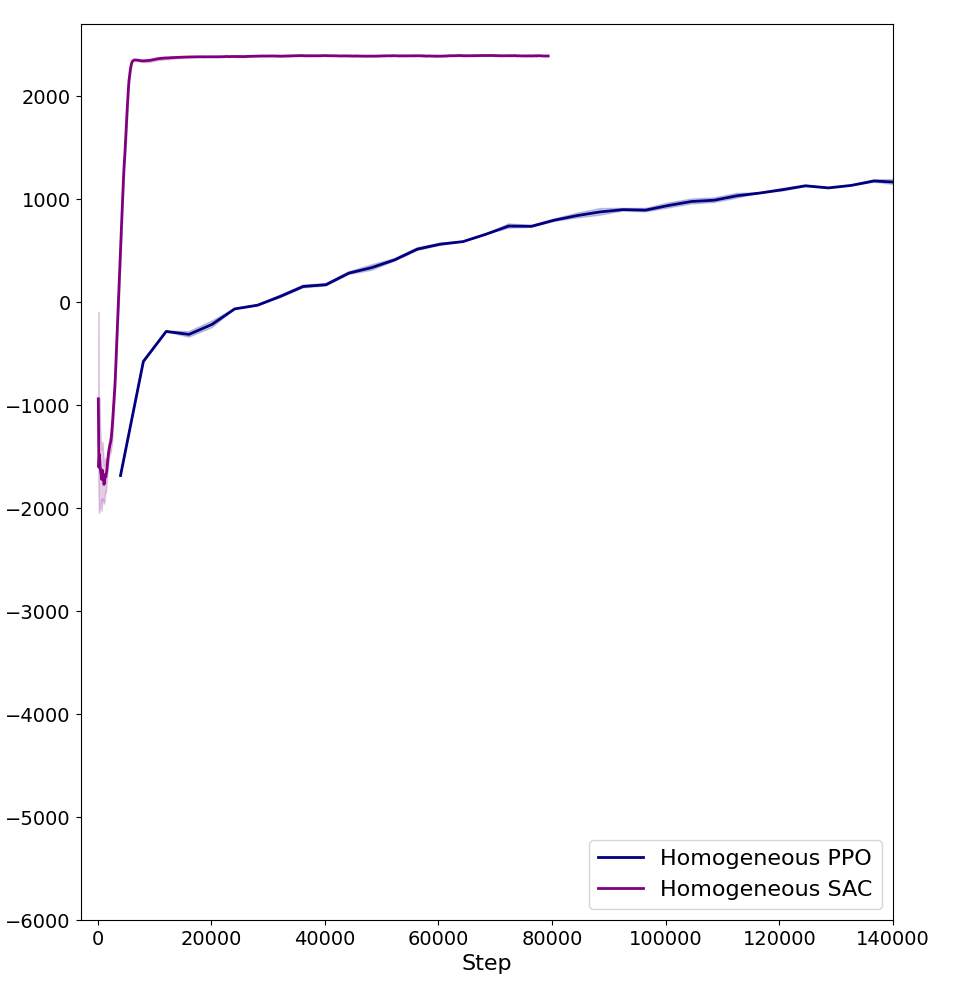}
\subcaption{Homogeneous agent .}
\end{subfigure}
\begin{subfigure}[!htbp]{0.485\linewidth}
        \centering
\includegraphics[width=\linewidth]{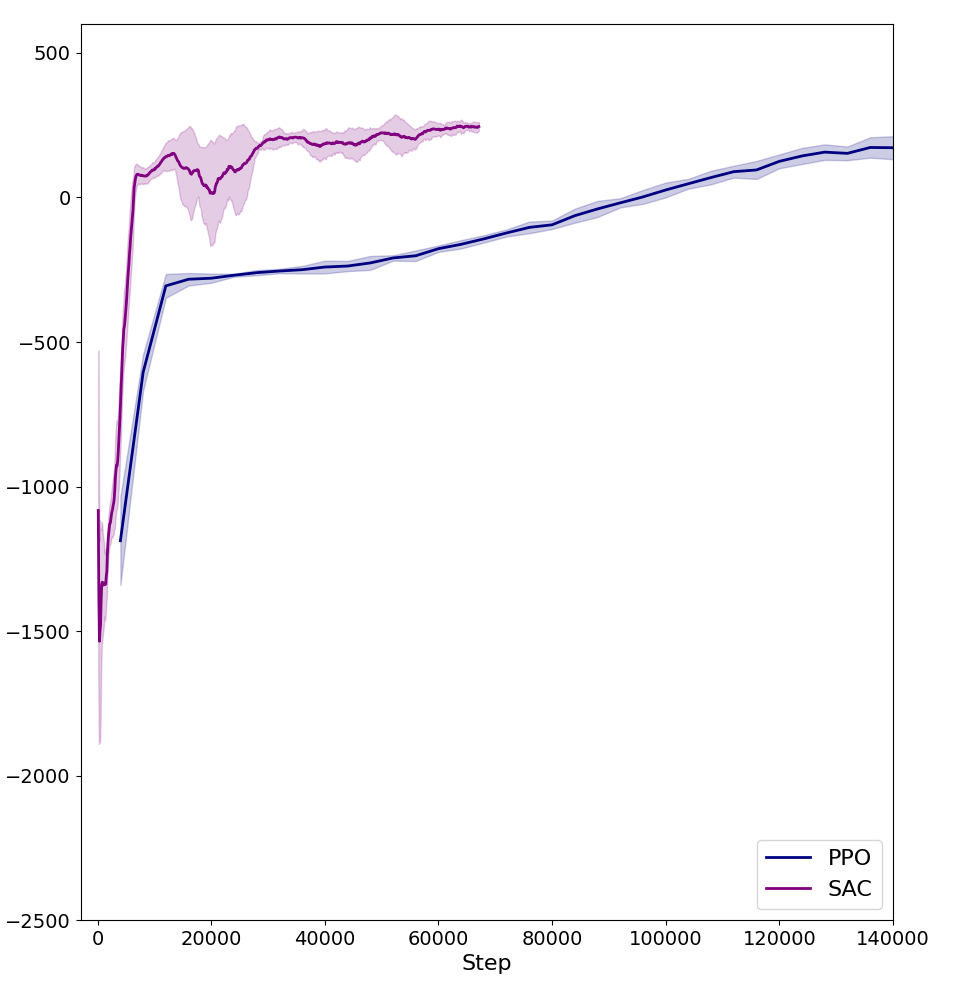}
\subcaption{Heterogeneous agent.}
\end{subfigure}
\caption{Homogeneous and Heterogeneous agent's performance in the low demand environment. The shaded area depicts the standard deviation of the multi-agent's performance for independent experiments using 5 different seeds.}
\label{fig:episode-reward-low}
\end{figure*}

The agents are generally less profitable in the low demand scenario compared to a high demand scenario with the Heterogeneous agents suffering a substantial drop in reward. Fig. \ref{fig:episode-reward-low} shows that in the low demand example the Heterogeneous agents receives lower rewards compared to the Homogeneous agents, with the SAC agent able to generate 224 and the PPO generating 208. However, the Homogeneous SAC agent generates 2,385 and the homogeneous PPO generates a slightly lower reward of 1,505.  

\begin{figure*}[!htbp]
    \centering
        \begin{subfigure}[!htbp]{0.46\linewidth}
        \centering
\includegraphics[width=\linewidth]{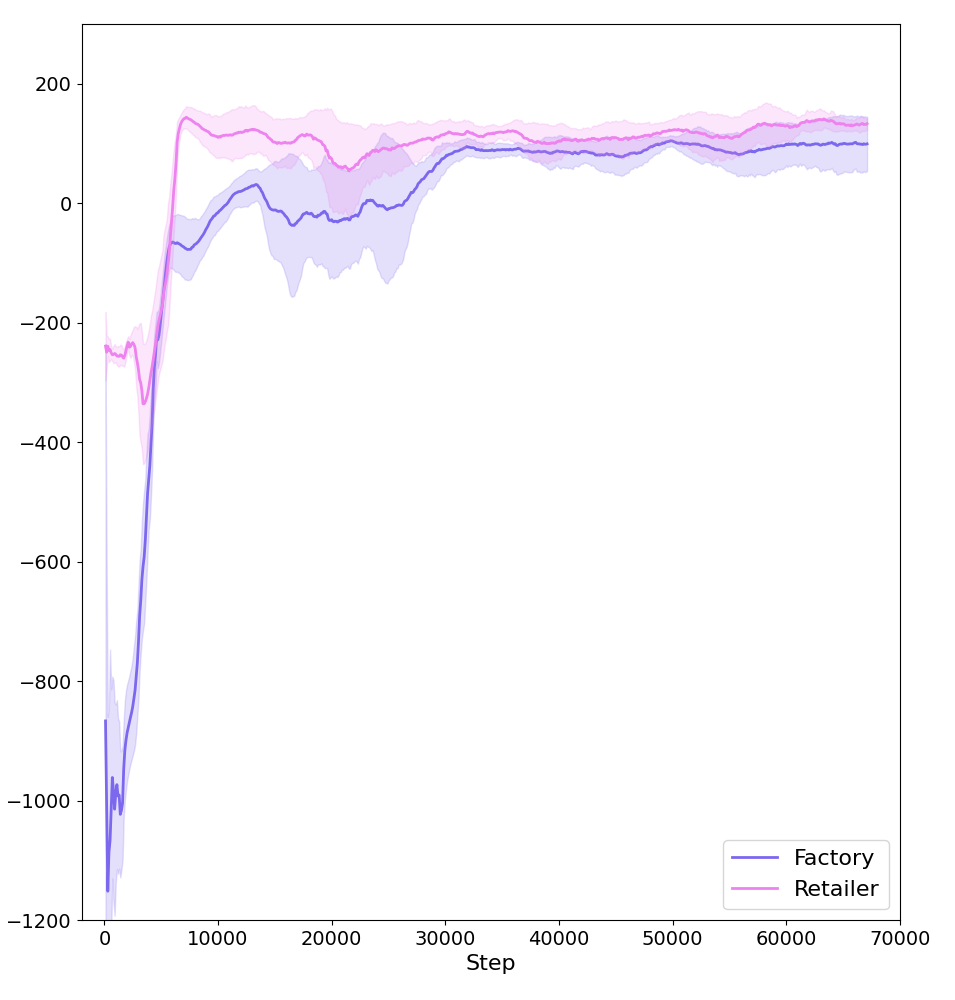}
  \subcaption{SAC}    \label{fig:episode}
  \end{subfigure}
    \begin{subfigure}[!htbp]{0.46\linewidth}
        \centering
\includegraphics[width=\linewidth]{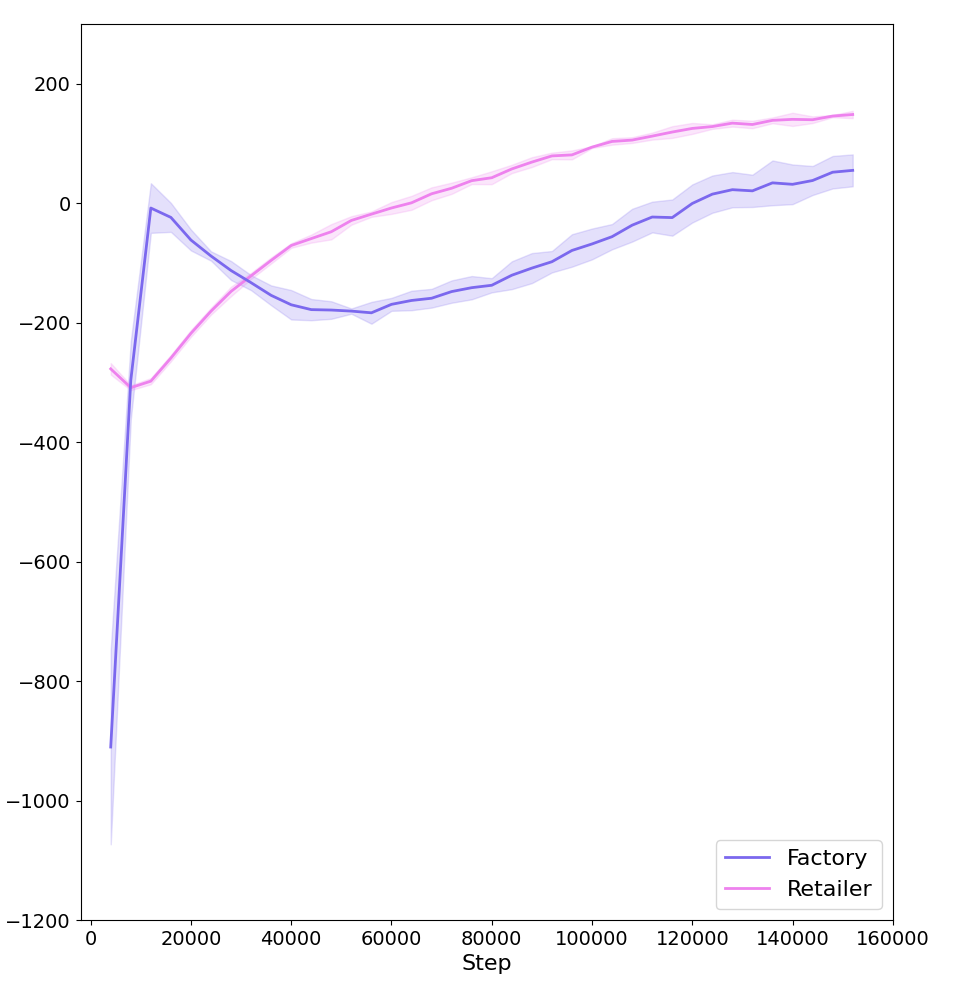}
  \subcaption{PPO}
\label{fig:episode}
  \end{subfigure}
\caption{Low demand heterogeneous agent's retailer and factory performance where the shaded area depicts the standard deviation of the independent experiments using 5 different seeds.}
\label{fig:factory-retailer-reward-low1}
\end{figure*}

Fig. \ref{fig:factory-retailer-reward-low1} demonstrates that for both the PPO and SAC algorithms the factory profit ends up lower than the retailer profit, reversing the trend shown in the high demand environment. For the PPO agent the retailer profit is 144 and the factory is 64 while for the SAC the Retailer is 131 and the factory generates a reward of 93. In these cases there is a higher variation in the performance for learners even in the converged stage of learning.

\begin{figure*}[!htbp]
        \includegraphics[width=\linewidth, height=6.6cm]{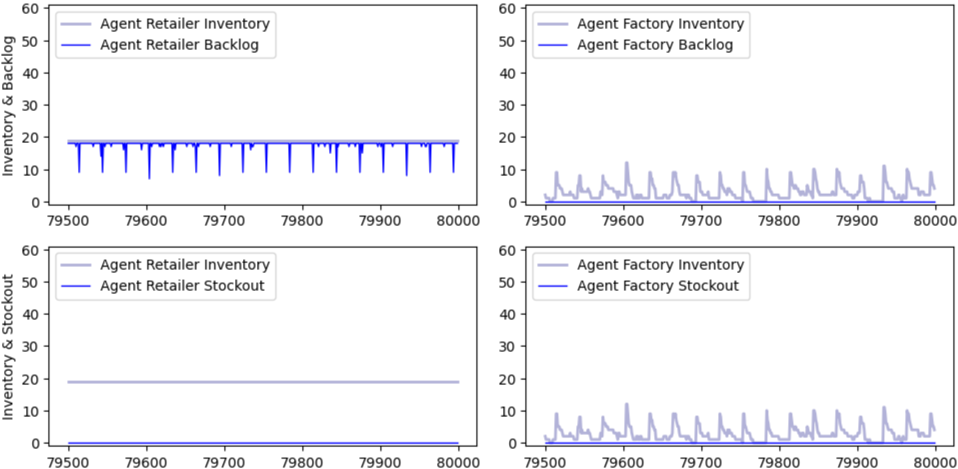} 
     \caption{Homogeneous SAC agent actions and resulting backlog and stockout in the low demand scenario. These methods have been proven to reduce the risk of stock-outs and backlogs in the supply chain by learning effective inventory strategies. }
     \label{fig:stockout-backlog-high2-homo}
\end{figure*}

In the low demand environment it is easier for the agent to avoid backlog and stockout. The ordering strategies show similarities to the high demand strategy but in this scenario it is easier for the agent to control the inventory. Similarly to the high demand environment the homogeneous SAC retailer agent keeps the stock at the highest level, with a mean of 19, despite the lower demand, demonstrated in Fig. \ref{fig:stockout-backlog-high2-homo}. This incurs a large number of backlog penalties but never incurs a stockout. The factory agent is able to respond with a regular inventory strategy, buying 9 or 10 items every few cycles and waiting for this stock to get used up. The factory overall inventory level remains low with a mean of 2.35. This incurs no stockout or backlog penalties. 

\begin{figure*}[!htbp]
        \centering
        \includegraphics[width=\linewidth, height=6.4cm]{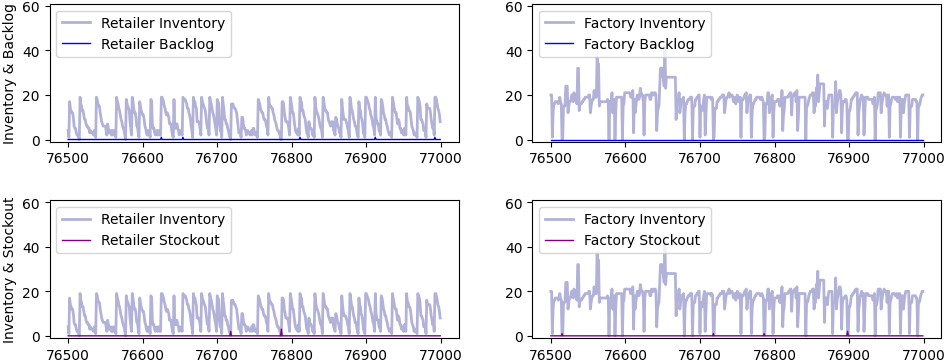} 
     \caption{Heterogeneous SAC agent actions and resulting backlog and stockout in the low demand scenario. These methods have been proven to reduce the risk of stock-outs and backlogs in the supply chain by learning effective inventory strategies.}
     \label{fig:stockout-backlog-low1-Hetero}
\end{figure*}

In the heterogeneous agent then there is a structured buying profile. The retailer agent buys enough stock to reach the maximum stock level and then lets this reduce to empty, shown in Fig. \ref{fig:stockout-backlog-low1-Hetero}, this leads to a mean of 8.52. The factory agent also keeps a higher stock level of 19.19 than in the homogeneous agent. In this case the agent tends to immediately recover it's stock level to the maximum as soon as possible, never allowing the stock to stay at a low level. It also has some erratic spikes where the inventory reaches levels close to 40. 

Similarly to the high demand scenario, the price for the different agents shows a clear relation to the performance. In this case the homogeneous agents charge the higher price with the SAC agent charging 5.7 and the PPO agent charging 5.27. This is substantially higher than the heterogeneous agents that charge lower amounts, with the heterogeneous PPO agent charging 3.69 and the heterogeneous SAC agent charging 3.08. In the heterogeneous cases, for both the PPO and the SAC agents, the variation across the range of different potential prices is high, there is no consistency with the agent sometimes selecting values in the 5-6 range but also selecting in the 0-1 range.


\begin{figure*}[!htbp]
    \centering
    \includegraphics[width=0.75\linewidth, height=6.3cm]{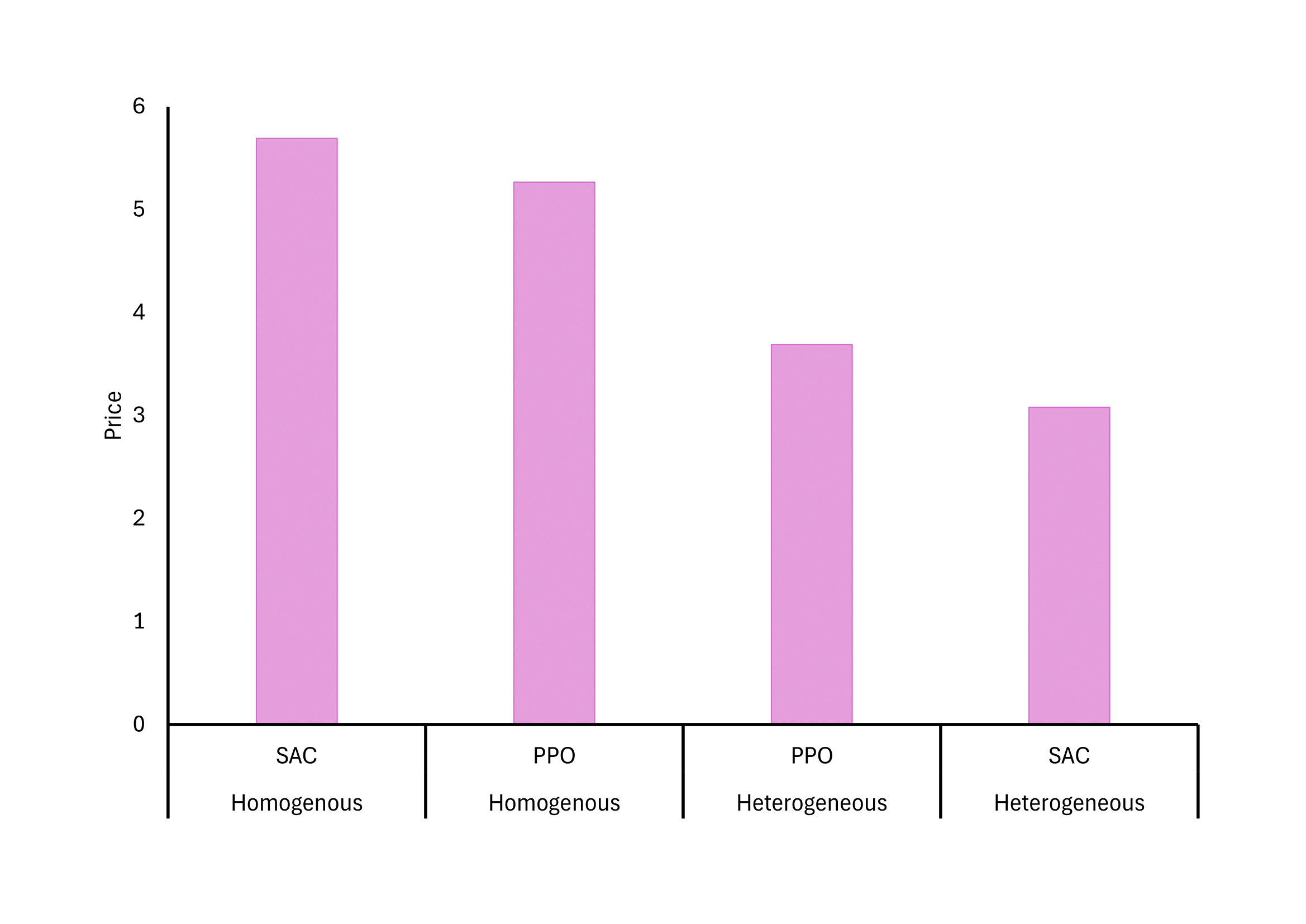} 
    \caption{Comparison of factory selling price across the different architectural agents in the low demand environment.}
    \label{fig:Low-Demand-Price}
\end{figure*}

\subsection{Reward shaping to increase Collaboration between agents in the low demand environment}
\label{Reward shaping low}
In the Collaborative reward structure the heterogeneous SAC agent has a maximum reward of 156 and the PPO agent receives 190 while the homogeneous SAC agent reaches rewards of 2,392 and the PPO agent reaches 1,229. 

\begin{figure*}[!htbp]
    \centering
     \begin{subfigure}[!htbp]{0.46\linewidth}
        \centering
\includegraphics[width=\linewidth]{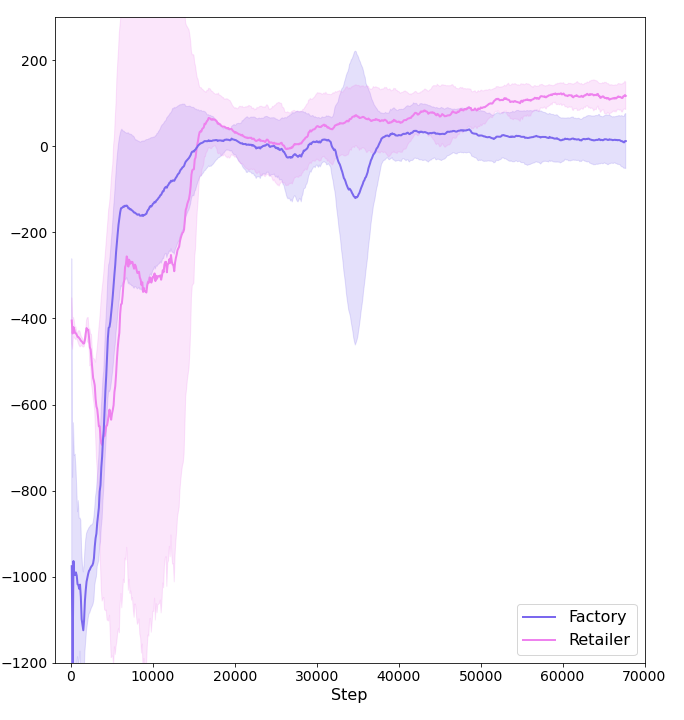}
  \subcaption{SAC in Collaborative}
\label{fig:episode}
  \end{subfigure}
   \begin{subfigure}[!htbp]{0.46\linewidth}
        \centering
\includegraphics[width=\linewidth]{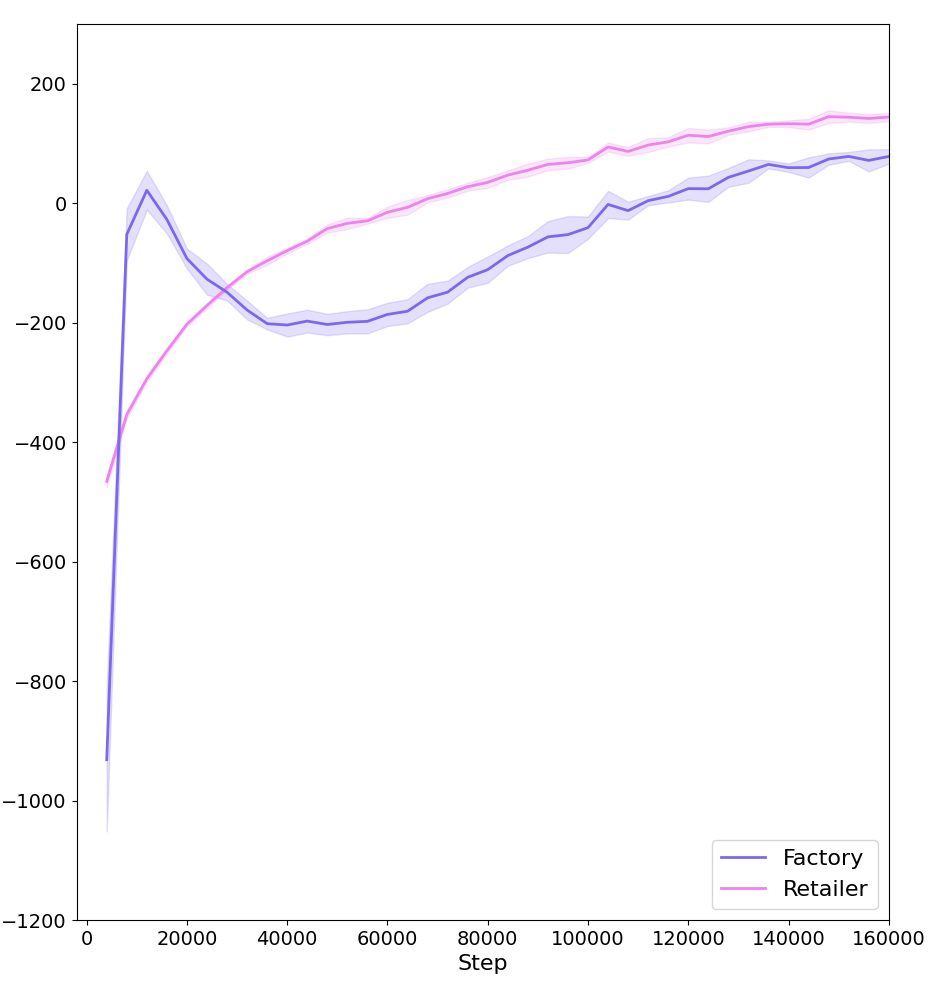}
  \subcaption{PPO in Collaborative}
\label{fig:episode}
  \end{subfigure}
\caption{Comparison of the factory and the retailer agent's reward for the SAC and PPO architectures in the low demand environment with reward sharing. The shaded area depicts the standard deviation of the heterogeneous agent's performance for independent experiments using 5 different seeds.}
\label{fig:factory-retailer-reward-low23}
\end{figure*}

Fig. \ref{fig:factory-retailer-reward-low23} provides the learning curves for the factory agent and retailer agents with collaborative reward sharing. The retailer agent again achieves higher profit.  For the SAC agent the collaborative retailer's profit is 127 while the collaborative factory agent's profit is 30.  PPO follows the same behaviour with the collaborative retailer agent getting a reward of 132  and the collaborative factory reward is 58. The factory reward is similar to that in the baseline environment, although the factory reward is lower. The SAC collaborative agent shows the highest variation in training for the retailer and the factory.

For the inventory, then the strategy remains the same as for the baseline environment. However, there are some small changes in the mean inventory, shown in Table \ref{Tab:collainv}. Here the factory generally has a lower inventory in the collaborative environment. This is the opposite of the high demand, where the baseline has a lower inventory.  However, the retailer is also lower in the homogeneous SAC agent which is the best performing. The low demand scenario shows a more general reduction in inventory across the 4 different architectures when the agents collaborate. 

\begin{table}[!htb]
     \centering
   \caption{Comparison of inventory levels of two symbiotic agents,  Benchmark and Collaborative scenarios, in a low demand supply chain.}
     \label{Tab:collainv}
    \begin{tabular}{p{2cm}p{1.5cm}p{2cm}p{2cm}p{2cm}p{2cm}}
         \toprule
      Architecture & Agent & Baseline Factory & Baseline Retailer& Collaboration Factory & Collaboration Retailer\\
        \midrule
Homogenous	&SAC	&19	& 2.35&\bf{18.998} &	\bf{1.836}\\
Homogeneous &	PPO	&	25.61 &	\bf{18.8}1&\bf{25.3}	&18.84\\
Heterogeneous	&PPO	&\bf{24.89}&	10.83&26.38	&\bf{9.965}\\
Heterogeneous	&SAC&		21.07 &	\bf{8.02}&\bf{19.7}&	8.36\\
         \bottomrule
    \end{tabular}
 \end{table}

The prices in the collaborative environments show limited differences to those in the standard environment. They are slightly higher except for the homogeneous PPO agent. The homogeneous SAC agent is 5.70 compared to 5.69 in the baseline, while the heterogeneous SAC agent has a mean of 3.15 compared to 3.08 for the baseline. For the PPO agents then the heterogeneous agent is 3.84 compared to 3.69 for the baseline and for the homogeneous agent the price is 5.05 compared to 5.27 for the baseline.  

The low demand environment leads to agents that perform differently to the high demand. In both the baseline and in the collaborative reward sharing the homogeneous agents outperform the heterogeneous agents. This is mainly related to the price, the homogenous agents are able to generate a strategy with a higher price. However, the control of the supply chain shifts significantly, with the retailer outperforming the factory. In this case, the reward sharing leads to more noticeable reductions in the inventory but no difference to the price being set.

\section{Discussion}
The demand changes control of the supply chain. In the high demand scenario, the factory controls the supply chain but in the low demand scenario it is the retailer. In the high demand environment, the performance is relatively even and this seems to relate to an even pricing strategy with all of the agents setting a high price. Here the type of agent is most important and SAC is the highest performer. This correlates with the performance of SAC in other RL environments, where the learner rapidly finds a near optimal policy \citep{Birkbeck2025}. 

In the low demand scenario, the high performance seems most related to being able to set a higher price. The homogeneous agents, with a single policy, are able to price significantly higher than the heterogeneous agents, which have separate policies. These values are repeatable, with the reward sharing results showing a similar final price. It is not clear whether the homogeneous agents set a higher price purely because of the observation space, but it also seems likely that this is related to the heterogeneous agents situation being a hidden mode Markov chain, which is significantly more challenging to solve and reduces the ability to collaborate. Heterogeneous configurations seem more likely in most real world scenarios, as nodes within the supply chain are likely to develop separate policies and this seems to play a larger role in the performance when the demand is low. 

\begin{figure*}[!htbp]
 \centering
   \begin{subfigure}[b]{0.47\linewidth}
        \centering
        \includegraphics[width=\linewidth, height=3cm]{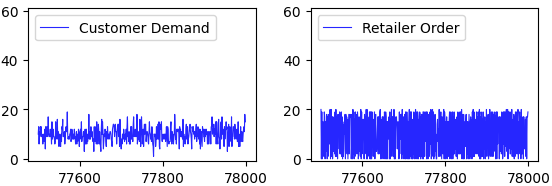} 
    \end{subfigure}
   \begin{subfigure}[b]{0.24\linewidth}
        \centering
        \includegraphics[width=\linewidth, height=3cm]{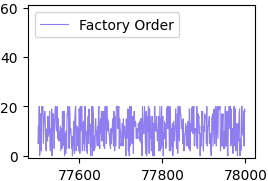} 
    \end{subfigure}
     \caption{Bullwhip effect and mitigation for the heterogeneous agents in high demand.}
     \label{fig:heter high}
\end{figure*}

The bullwhip effect occurs when a lag in demand forecasts causes growing oscillations in inventory levels, analogous to the motion of a whip\citep{lee1997bullwhip}. The agent architecture and demand are the key indicators for behaviour. The two heterogeneous agents, PPO and SAC, show similar behaviours with a change in the demand creating a change in strategy. In the high demand scenario, the bullwhip effect from customer to retailer is high, shown in Fig. \ref{fig:heter high}. However, this is mitigated from the retailer to the factory, with less extreme fluctuations.  

\begin{figure*}[!htbp]
 \centering
   \begin{subfigure}[b]{0.47\linewidth}
        \centering
        \includegraphics[width=\linewidth, height=3cm]{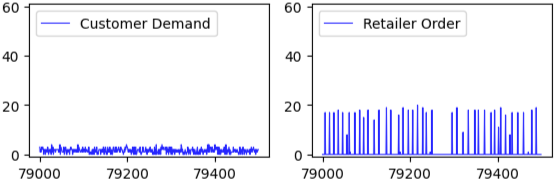} 
    \end{subfigure}
   \begin{subfigure}[b]{0.24\linewidth}
        \centering
        \includegraphics[width=\linewidth, height=3cm]{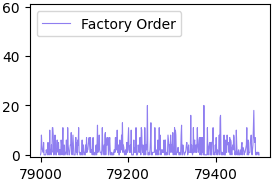} 
    \end{subfigure}
     \caption{Bullwhip effect and mitigation for the heterogeneous agents in low demand.  }
     \label{fig:heter low}
\end{figure*}

The behaviour for the low demand scenario is more extreme, with the retailer making regular large orders and then waiting for the stock to dissipate, shown in Fig. \ref{fig:heter low}. This becomes less regular in the factory, with the peaks of ordering being more stochastic and with more regular ordering in between. If there was a longer chain, the heterogeneous agents should be able to effectively mitigate the bullwhip effect. 

\begin{figure*}[!htbp]
 \centering
   \begin{subfigure}[b]{0.78\linewidth}
        \centering
        \includegraphics[width=\linewidth, height=3cm]{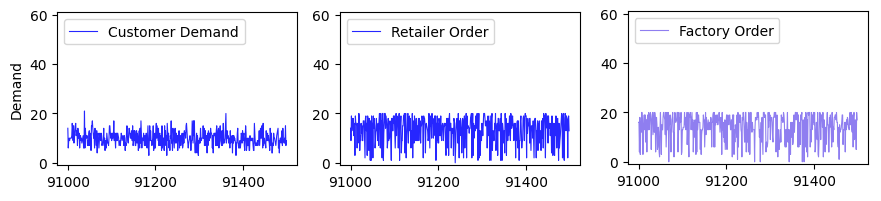} 
    \end{subfigure}
     \caption{Bullwhip effect for the PPO Homogenous Agents.  }
     \label{fig:homo PPO}
\end{figure*}

The homogeneous agents show a difference based on the type of agent, SAC or PPO, but the strategies stay consistent across the demands. This leads to large numbers of backlogs. Fig. \ref{fig:homo PPO} shows that the PPO agent retains a relatively regular buying strategy, at almost the maximum orders. This is passed on to the factory, which exhibits a similar buying strategy. 

\begin{figure*}[!htbp]
 \centering
   \begin{subfigure}[b]{0.78\linewidth}
        \centering
        \includegraphics[width=\linewidth, height=3cm]{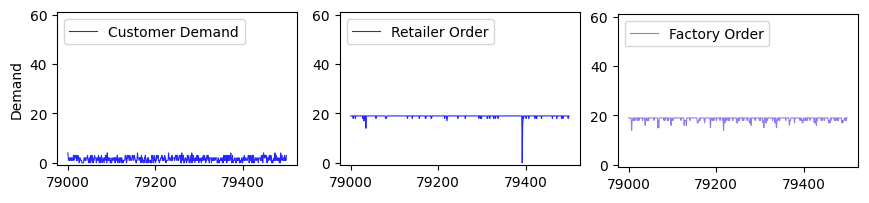} 
    \end{subfigure}
     \caption{Bullwhip effect for the SAC Homogenous Agents.  }
     \label{fig:homo SAC}
\end{figure*}

This is replicated in the SAC agent with an even more consistent buying strategy, always maximising the inventory that it buys as shown in Fig. \ref{fig:homo SAC}. This is replicated in the factory, which needs to buy the maximum amount of stock each round. In these cases the heterogeneous agents are able to mitigate the bullwhip effect, but the homogenous agents have such high buying strategies in both environments that it is passed between the echelons. 

A number of these strategies look similar to those considered in fundamental theory. Inventories are considered to fluctuate from a maximum when the order is made through a linear decrease to 0 and are then replenished. The homogeneous factory agent clearly shows this pattern for the low and high demand experiments, as shown Figure \ref{fig:stockout-backlog-high2} and Figure \ref{fig:stockout-backlog-high2-homo}. Economic Order Quantity (EOQ) is a simplified calculation to help determine the order size, given in Eq. \ref{E:EOQ},

\begin{equation}
\label{E:EOQ}
Q^* = \sqrt{\frac{2D \times Oc}{Hc}} \times \sqrt{\frac{Hc + Sc}{Sc}},
\end{equation}

$Q^*$ signifies the Economic Order Quantity, $D$ represents the demand in units,  $Oc$ represents order cost(such as transportation, setup, or administrative cost) each time,  $Sc$ represents stockout cost and  $Hc$ reflects unit holding cost. In the high demand scenario, the factory mean inventory is 18.91 for the homogeneous SAC, this is close to 32\% of the inventory capacity, and 30.71 for the heterogeneous SAC, which is 52\%. For the retailer the mean inventory is 3.12 for the homogeneous SAC, which is 16\% of the inventory capacity, and 10.41 for the heterogeneous SAC, which is 55\%. In the low demand experiments the factory mean inventory is 19 for the homogeneous SAC, which is 32\%, and 21.07 for the heterogeneous SAC which is 35\%. For the retailer the mean inventory is 2.35 for the homogeneous SAC, which is close to 12\% of the inventory capacity, and 8.02 for the heterogeneous SAC, which is 42\%. It is found that the retailer order quantity of the heterogeneous agent algorithm is closer to the Economic Order Quantity (EOQ) than the homogeneous agent algorithm.


Reward sharing without sharing profit is shown to have an effect on the agent behaviour. The trend reverses between the low and high demand, with the retailer generally showing a lower inventory in the high demand and the factory generally taking on the lower inventory in the low demand.
These trends aren't totally consistent across the different architectures and agents. Greater observability of the environment might need to be considered to allow the agents to be able to work together to a higher extent. In the high demand environment, it seems that it is difficult to find a strategy that can fulfil the demand and there is limited opportunity to adapt the strategy to allow for greater cooperation. In the low demand environment, it is easy to ensure that the agent avoids stockouts and there is no need for this type of strategy. There is perhaps a Goldilocks demand, where this reward sharing becomes most effective but it seems unrealistic to tune the sharing to this extent with most demands varying over time. In the same manner, the penalty for stockouts could be increased but this seems to be unrealisitic. While most supply chains will not consider a profit share, as implemented in the previous literature, \cite{liu2022multi} and \cite{yu2020multi}, other strategies need to be considered, perhaps integrating the ability for firms to go bust if they are performing poorly or through understanding whether great communication about the environment allows more complex cooperative environments to be developed.

\section{Conclusion}
Supply Chain Management (SCM) involves coordinating the flow of goods, information, and money between different entities to deliver products efficiently. Multi-agent algorithms are being explored to control this flow by allowing agents to learn how to develop their own strategies based on the actions of others. In the literature two architecture types are explored, single policy, homogeneous, and multiple policy approaches, heterogeneous. This changes the observability of the problem, with homogeneous agents deemed to be less realistic for most supply chains as they will require an exceptional level of trust to implement. This paper investigates how heterogeneous and homogeneous agents perform in comparison to each other. Two environments are explored, a high demand and a low demand environment, alongside two algorithms, PPO and SAC. Reward sharing without sharing profit is found to be difficult, showing a small change in behaviour based on this change. With both reward sharing and a vanilla reward, the homogeneous and heterogeneous agents exhibit different behaviours, with the homogeneous retailer retaining high inventories and witnessing a lot of backlog and heterogeneous agents retaining a lower inventory. This leads to the heterogeneous agents mitigating the bullwhip effect but this is passed on in the homogeneous supply chains. In the high demand environment, the agent architecture dominates performance with the SAC agents outperforming the PPO agents. In this environment the factory controls the supply chain. In the low demand environment, control of the supply chain shifts significantly, with the retailer outperforming the factory by a significant margin.  In these condition, the homogenous agents outperform the heterogeneous agents. Indicating that some of the literature might be optimisitic about the capabilities of current multi-agent systems. The performance is mainly related to charging higher prices, which is inhibited by the Hidden Markov Process when heterogeneous agents are implemented.

\section*{CRediT}
Wan Wang Conceptualization; Funding acquisition; Investigation; Methodology; Project administration; Validation; Visualization; Roles/Writing - original draft; and Writing - review \& editing
Haiyan  Funding acquisition; Project administration; Supervision; Writing - review \& editing
Adam Conceptualization; Methodology; Supervision; and Writing - review \& editing

\section*{Acknowledgments}
The authors acknowledge the use of the IRIDIS High-Performance Computing Facility, and associated support services at the University of Southampton, in the completion of this work. Wan Wang is funded by the China Scholarship Council. We would like to thank Lloyd’s Register Foundation for their support during this research. Any errors or discrepancies are our own responsibility.

\label{reference_style}
\bibliography{arxiv}

\begin{thebibliography}{37}
\expandafter\ifx\csname natexlab\endcsname\relax\def\natexlab#1{#1}\fi
\providecommand{\url}[1]{\texttt{#1}}
\providecommand{\href}[2]{#2}
\providecommand{\path}[1]{#1}
\providecommand{\DOIprefix}{doi:}
\providecommand{\ArXivprefix}{arXiv:}
\providecommand{\URLprefix}{URL: }
\providecommand{\Pubmedprefix}{pmid:}
\providecommand{\doi}[1]{\href{http://dx.doi.org/#1}{\path{#1}}}
\providecommand{\Pubmed}[1]{\href{pmid:#1}{\path{#1}}}
\providecommand{\bibinfo}[2]{#2}
\ifx\xfnm\relax \def\xfnm[#1]{\unskip,\space#1}\fi
\bibitem[{Alamdar \& Seifi(2024)}]{alamdar2024deep}
\bibinfo{author}{Alamdar, P.~F.}, \& \bibinfo{author}{Seifi, A.} (\bibinfo{year}{2024}).
\newblock \bibinfo{title}{A deep q-learning approach to optimize ordering and dynamic pricing decisions in the presence of strategic customers}.
\newblock {\it \bibinfo{journal}{International Journal of Production Economics}\/},  {\it \bibinfo{volume}{269}\/}, \bibinfo{pages}{109154}.
\bibitem[{Arifo{\u{g}}lu \& {\"O}zekici(2010)}]{arifouglu2010optimal}
\bibinfo{author}{Arifo{\u{g}}lu, K.}, \& \bibinfo{author}{{\"O}zekici, S.} (\bibinfo{year}{2010}).
\newblock \bibinfo{title}{Optimal policies for inventory systems with finite capacity and partially observed markov-modulated demand and supply processes}.
\newblock {\it \bibinfo{journal}{European Journal of Operational Research}\/},  {\it \bibinfo{volume}{204}\/}, \bibinfo{pages}{421--438}.
\bibitem[{Birkbeck et~al.(2024)Birkbeck, Sobey, Cerutti, Heseltine Hurley~Flynn \& Norman}]{Birkbeck2025}
\bibinfo{author}{Birkbeck, J.}, \bibinfo{author}{Sobey, A.}, \bibinfo{author}{Cerutti, F.}, \bibinfo{author}{Heseltine Hurley~Flynn, K.}, \& \bibinfo{author}{Norman, T.} (\bibinfo{year}{2024}).
\newblock \bibinfo{title}{Chirps: Change-induced regret proxy metrics for lifelong reinforcement learning}.
\newblock {\it \bibinfo{journal}{https://arxiv.org/abs/2409.03577}\/}, .
\bibitem[{Brintrup(2010)}]{brintrup2010behaviour}
\bibinfo{author}{Brintrup, A.} (\bibinfo{year}{2010}).
\newblock \bibinfo{title}{Behaviour adaptation in the multi-agent, multi-objective and multi-role supply chain}.
\newblock {\it \bibinfo{journal}{Computers in Industry}\/},  {\it \bibinfo{volume}{61}\/}, \bibinfo{pages}{636--645}.
\bibitem[{Brockman et~al.(2016)Brockman, Cheung, Pettersson, Schneider, Schulman, Tang \& Zaremba}]{brockman2016openai}
\bibinfo{author}{Brockman, G.}, \bibinfo{author}{Cheung, V.}, \bibinfo{author}{Pettersson, L.}, \bibinfo{author}{Schneider, J.}, \bibinfo{author}{Schulman, J.}, \bibinfo{author}{Tang, J.}, \& \bibinfo{author}{Zaremba, W.} (\bibinfo{year}{2016}).
\newblock \bibinfo{title}{Openai gym}.
\newblock {\it \bibinfo{journal}{arXiv preprint arXiv:1606.01540}\/}, .
\bibitem[{Chen(2021)}]{chen2021data}
\bibinfo{author}{Chen, B.} (\bibinfo{year}{2021}).
\newblock \bibinfo{title}{Data-driven inventory control with shifting demand}.
\newblock {\it \bibinfo{journal}{Production and Operations Management}\/},  {\it \bibinfo{volume}{30}\/}, \bibinfo{pages}{1365--1385}.
\bibitem[{Ding et~al.(2022)Ding, Feng, Liu, Jiang, Zhang, Zhao, Song, Li, Jin \& Bian}]{ding2022multi}
\bibinfo{author}{Ding, Y.}, \bibinfo{author}{Feng, M.}, \bibinfo{author}{Liu, G.}, \bibinfo{author}{Jiang, W.}, \bibinfo{author}{Zhang, C.}, \bibinfo{author}{Zhao, L.}, \bibinfo{author}{Song, L.}, \bibinfo{author}{Li, H.}, \bibinfo{author}{Jin, Y.}, \& \bibinfo{author}{Bian, J.} (\bibinfo{year}{2022}).
\newblock \bibinfo{title}{Multi-agent reinforcement learning with shared resources for inventory management}.
\newblock {\it \bibinfo{journal}{arXiv preprint arXiv:2212.07684}\/}, .
\bibitem[{Dogan \& G{\"u}ner(2015)}]{dogan2015reinforcement}
\bibinfo{author}{Dogan, I.}, \& \bibinfo{author}{G{\"u}ner, A.~R.} (\bibinfo{year}{2015}).
\newblock \bibinfo{title}{A reinforcement learning approach to competitive ordering and pricing problem}.
\newblock {\it \bibinfo{journal}{Expert Systems}\/},  {\it \bibinfo{volume}{32}\/}, \bibinfo{pages}{39--48}.
\bibitem[{Gijsbrechts et~al.(2022)Gijsbrechts, Boute, Van~Mieghem \& Zhang}]{gijsbrechts2022can}
\bibinfo{author}{Gijsbrechts, J.}, \bibinfo{author}{Boute, R.~N.}, \bibinfo{author}{Van~Mieghem, J.~A.}, \& \bibinfo{author}{Zhang, D.~J.} (\bibinfo{year}{2022}).
\newblock \bibinfo{title}{Can deep reinforcement learning improve inventory management? performance on lost sales, dual-sourcing, and multi-echelon problems}.
\newblock {\it \bibinfo{journal}{Manufacturing \& Service Operations Management}\/},  {\it \bibinfo{volume}{24}\/}, \bibinfo{pages}{1349--1368}.
\bibitem[{Guo et~al.(2023)Guo, Chen, Boulaksil, Xiao \& Allaoui}]{guo2023collaborative}
\bibinfo{author}{Guo, Y.}, \bibinfo{author}{Chen, T.}, \bibinfo{author}{Boulaksil, Y.}, \bibinfo{author}{Xiao, L.}, \& \bibinfo{author}{Allaoui, H.} (\bibinfo{year}{2023}).
\newblock \bibinfo{title}{Collaborative planning of multi-tier sustainable supply chains: A reinforcement learning enhanced heuristic approach}.
\newblock {\it \bibinfo{journal}{Computers \& Industrial Engineering}\/},  {\it \bibinfo{volume}{185}\/}, \bibinfo{pages}{109669}.
\bibitem[{Hekimo{\u{g}}lu et~al.(2018)Hekimo{\u{g}}lu, van~der Laan \& Dekker}]{hekimouglu2018markov}
\bibinfo{author}{Hekimo{\u{g}}lu, M.}, \bibinfo{author}{van~der Laan, E.}, \& \bibinfo{author}{Dekker, R.} (\bibinfo{year}{2018}).
\newblock \bibinfo{title}{Markov-modulated analysis of a spare parts system with random lead times and disruption risks}.
\newblock {\it \bibinfo{journal}{European Journal of Operational Research}\/},  {\it \bibinfo{volume}{269}\/}, \bibinfo{pages}{909--922}.
\bibitem[{Hubbs et~al.(2020)Hubbs, Perez, Sarwar, Sahinidis, Grossmann \& Wassick}]{hubbs2020or}
\bibinfo{author}{Hubbs, C.~D.}, \bibinfo{author}{Perez, H.~D.}, \bibinfo{author}{Sarwar, O.}, \bibinfo{author}{Sahinidis, N.~V.}, \bibinfo{author}{Grossmann, I.~E.}, \& \bibinfo{author}{Wassick, J.~M.} (\bibinfo{year}{2020}).
\newblock \bibinfo{title}{Or-gym: A reinforcement learning library for operations research problems}.
\newblock {\it \bibinfo{journal}{arXiv preprint arXiv:2008.06319}\/}, .
\bibitem[{Jiang et~al.(2023)Jiang, Hu \& Peng}]{jiang2023quantile}
\bibinfo{author}{Jiang, J.}, \bibinfo{author}{Hu, J.}, \& \bibinfo{author}{Peng, Y.} (\bibinfo{year}{2023}).
\newblock \bibinfo{title}{Quantile-based deep reinforcement learning using two-timescale policy gradient algorithms}.
\newblock {\it \bibinfo{journal}{arXiv preprint arXiv:2305.07248}\/}, .
\bibitem[{Keskin et~al.(2022)Keskin, Li \& Song}]{keskin2022data}
\bibinfo{author}{Keskin, N.~B.}, \bibinfo{author}{Li, Y.}, \& \bibinfo{author}{Song, J.-S.} (\bibinfo{year}{2022}).
\newblock \bibinfo{title}{Data-driven dynamic pricing and ordering with perishable inventory in a changing environment}.
\newblock {\it \bibinfo{journal}{Management Science}\/},  {\it \bibinfo{volume}{68}\/}, \bibinfo{pages}{1938--1958}.
\bibitem[{Kim et~al.(2024)Kim, Kim \& Lee}]{kim2024multi}
\bibinfo{author}{Kim, B.}, \bibinfo{author}{Kim, J.~G.}, \& \bibinfo{author}{Lee, S.} (\bibinfo{year}{2024}).
\newblock \bibinfo{title}{A multi-agent reinforcement learning model for inventory transshipments under supply chain disruption}.
\newblock {\it \bibinfo{journal}{IISE Transactions}\/},  {\it \bibinfo{volume}{56}\/}, \bibinfo{pages}{715--728}.
\bibitem[{Kosasih \& Brintrup(2022)}]{kosasih2022reinforcement}
\bibinfo{author}{Kosasih, E.~E.}, \& \bibinfo{author}{Brintrup, A.} (\bibinfo{year}{2022}).
\newblock \bibinfo{title}{Reinforcement learning provides a flexible approach for realistic supply chain safety stock optimisation}.
\newblock {\it \bibinfo{journal}{IFAC-PapersOnLine}\/},  {\it \bibinfo{volume}{55}\/}, \bibinfo{pages}{1539--1544}.
\bibitem[{Lee et~al.(1997)Lee, Padmanabhan \& Whang}]{lee1997bullwhip}
\bibinfo{author}{Lee, H.~L.}, \bibinfo{author}{Padmanabhan, V.}, \& \bibinfo{author}{Whang, S.} (\bibinfo{year}{1997}).
\newblock \bibinfo{title}{The bullwhip effect in supply chains}, .
\bibitem[{Leluc et~al.(2023)Leluc, Kadoche, Bertoncello \& Gourv{\'e}nec}]{leluc2023marlim}
\bibinfo{author}{Leluc, R.}, \bibinfo{author}{Kadoche, E.}, \bibinfo{author}{Bertoncello, A.}, \& \bibinfo{author}{Gourv{\'e}nec, S.} (\bibinfo{year}{2023}).
\newblock \bibinfo{title}{Marlim: Multi-agent reinforcement learning for inventory management}.
\newblock {\it \bibinfo{journal}{arXiv preprint arXiv:2308.01649}\/}, .
\bibitem[{Liang et~al.(2018)Liang, Liaw, Nishihara, Moritz, Fox, Goldberg, Gonzalez, Jordan \& Stoica}]{liang2018rllib}
\bibinfo{author}{Liang, E.}, \bibinfo{author}{Liaw, R.}, \bibinfo{author}{Nishihara, R.}, \bibinfo{author}{Moritz, P.}, \bibinfo{author}{Fox, R.}, \bibinfo{author}{Goldberg, K.}, \bibinfo{author}{Gonzalez, J.}, \bibinfo{author}{Jordan, M.}, \& \bibinfo{author}{Stoica, I.} (\bibinfo{year}{2018}).
\newblock \bibinfo{title}{Rllib: Abstractions for distributed reinforcement learning}.
\newblock In {\it \bibinfo{booktitle}{International conference on machine learning}\/} (pp. \bibinfo{pages}{3053--3062}).
\newblock \bibinfo{organization}{PMLR}.
\bibitem[{Liu et~al.(2022)Liu, Hu, Peng \& Yang}]{liu2022multi}
\bibinfo{author}{Liu, X.}, \bibinfo{author}{Hu, M.}, \bibinfo{author}{Peng, Y.}, \& \bibinfo{author}{Yang, Y.} (\bibinfo{year}{2022}).
\newblock \bibinfo{title}{Multi-agent deep reinforcement learning for multi-echelon inventory management}.
\newblock {\it \bibinfo{journal}{Available at SSRN}\/}, .
\bibitem[{Moritz et~al.(2018)Moritz, Nishihara, Wang, Tumanov, Liaw, Liang, Elibol, Yang, Paul, Jordan et~al.}]{moritz2018ray}
\bibinfo{author}{Moritz, P.}, \bibinfo{author}{Nishihara, R.}, \bibinfo{author}{Wang, S.}, \bibinfo{author}{Tumanov, A.}, \bibinfo{author}{Liaw, R.}, \bibinfo{author}{Liang, E.}, \bibinfo{author}{Elibol, M.}, \bibinfo{author}{Yang, Z.}, \bibinfo{author}{Paul, W.}, \bibinfo{author}{Jordan, M.~I.} et~al. (\bibinfo{year}{2018}).
\newblock \bibinfo{title}{Ray: A distributed framework for emerging $\{$AI$\}$ applications}.
\newblock In {\it \bibinfo{booktitle}{13th USENIX symposium on operating systems design and implementation (OSDI 18)}\/} (pp. \bibinfo{pages}{561--577}).
\bibitem[{Mousa et~al.(2024)Mousa, van~de Berg, Kotecha, del Rio~Chanona \& Mowbray}]{mousa2024analysis}
\bibinfo{author}{Mousa, M.}, \bibinfo{author}{van~de Berg, D.}, \bibinfo{author}{Kotecha, N.}, \bibinfo{author}{del Rio~Chanona, E.~A.}, \& \bibinfo{author}{Mowbray, M.} (\bibinfo{year}{2024}).
\newblock \bibinfo{title}{An analysis of multi-agent reinforcement learning for decentralized inventory control systems}.
\newblock {\it \bibinfo{journal}{Computers \& Chemical Engineering}\/},  {\it \bibinfo{volume}{188}\/}, \bibinfo{pages}{108783}.
\bibitem[{Oroojlooyjadid et~al.(2022)Oroojlooyjadid, Nazari, Snyder \& Tak{\'a}{\v{c}}}]{oroojlooyjadid2022deep}
\bibinfo{author}{Oroojlooyjadid, A.}, \bibinfo{author}{Nazari, M.}, \bibinfo{author}{Snyder, L.~V.}, \& \bibinfo{author}{Tak{\'a}{\v{c}}, M.} (\bibinfo{year}{2022}).
\newblock \bibinfo{title}{A deep q-network for the beer game: Deep reinforcement learning for inventory optimization}.
\newblock {\it \bibinfo{journal}{Manufacturing \& Service Operations Management}\/},  {\it \bibinfo{volume}{24}\/}, \bibinfo{pages}{285--304}.
\bibitem[{Paine(2022)}]{paine2022behaviorally}
\bibinfo{author}{Paine, J.} (\bibinfo{year}{2022}).
\newblock \bibinfo{title}{Behaviorally grounded model-based and model free cost reduction in a simulated multi-echelon supply chain}.
\newblock {\it \bibinfo{journal}{arXiv preprint arXiv:2202.12786}\/}, .
\bibitem[{Qiao et~al.(2024)Qiao, Huang, Gao \& Wang}]{qiao2024distributed}
\bibinfo{author}{Qiao, W.}, \bibinfo{author}{Huang, M.}, \bibinfo{author}{Gao, Z.}, \& \bibinfo{author}{Wang, X.} (\bibinfo{year}{2024}).
\newblock \bibinfo{title}{Distributed dynamic pricing of multiple perishable products using multi-agent reinforcement learning}.
\newblock {\it \bibinfo{journal}{Expert Systems with Applications}\/},  {\it \bibinfo{volume}{237}\/}, \bibinfo{pages}{121252}.
\bibitem[{Shi et~al.(2016)Shi, Chen \& Duenyas}]{shi2016nonparametric}
\bibinfo{author}{Shi, C.}, \bibinfo{author}{Chen, W.}, \& \bibinfo{author}{Duenyas, I.} (\bibinfo{year}{2016}).
\newblock \bibinfo{title}{Nonparametric data-driven algorithms for multiproduct inventory systems with censored demand}.
\newblock {\it \bibinfo{journal}{Operations Research}\/},  {\it \bibinfo{volume}{64}\/}, \bibinfo{pages}{362--370}.
\bibitem[{Stranieri \& Stella(2022)}]{stranieri2022deep}
\bibinfo{author}{Stranieri, F.}, \& \bibinfo{author}{Stella, F.} (\bibinfo{year}{2022}).
\newblock \bibinfo{title}{A deep reinforcement learning approach to supply chain inventory management}.
\newblock {\it \bibinfo{journal}{arXiv preprint arXiv:2204.09603}\/}, .
\bibitem[{Stranieri et~al.(2024)Stranieri, Stella \& Kouki}]{stranieri2024performance}
\bibinfo{author}{Stranieri, F.}, \bibinfo{author}{Stella, F.}, \& \bibinfo{author}{Kouki, C.} (\bibinfo{year}{2024}).
\newblock \bibinfo{title}{Performance of deep reinforcement learning algorithms in two-echelon inventory control systems}.
\newblock {\it \bibinfo{journal}{International Journal of Production Research}\/},  (pp. \bibinfo{pages}{1--16}).
\bibitem[{Sultana et~al.(2020)Sultana, Meisheri, Baniwal, Nath, Ravindran \& Khadilkar}]{sultana2020reinforcement}
\bibinfo{author}{Sultana, N.~N.}, \bibinfo{author}{Meisheri, H.}, \bibinfo{author}{Baniwal, V.}, \bibinfo{author}{Nath, S.}, \bibinfo{author}{Ravindran, B.}, \& \bibinfo{author}{Khadilkar, H.} (\bibinfo{year}{2020}).
\newblock \bibinfo{title}{Reinforcement learning for multi-product multi-node inventory management in supply chains}.
\newblock {\it \bibinfo{journal}{arXiv preprint arXiv:2006.04037}\/}, .
\bibitem[{Tian et~al.(2024)Tian, Lu, Wang, Wang \& Tang}]{tian2024iacppo}
\bibinfo{author}{Tian, R.}, \bibinfo{author}{Lu, M.}, \bibinfo{author}{Wang, H.}, \bibinfo{author}{Wang, B.}, \& \bibinfo{author}{Tang, Q.} (\bibinfo{year}{2024}).
\newblock \bibinfo{title}{Iacppo: A deep reinforcement learning-based model for warehouse inventory replenishment}.
\newblock {\it \bibinfo{journal}{Computers \& Industrial Engineering}\/},  {\it \bibinfo{volume}{187}\/}, \bibinfo{pages}{109829}.
\bibitem[{Toomey(2000)}]{toomey2000inventory}
\bibinfo{author}{Toomey, J.~W.} (\bibinfo{year}{2000}).
\newblock {\it \bibinfo{title}{Inventory management: principles, concepts and techniques}\/} volume~\bibinfo{volume}{12}.
\newblock \bibinfo{publisher}{Springer Science \& Business Media}.
\bibitem[{Vanvuchelen et~al.(2020)Vanvuchelen, Gijsbrechts \& Boute}]{vanvuchelen2020use}
\bibinfo{author}{Vanvuchelen, N.}, \bibinfo{author}{Gijsbrechts, J.}, \& \bibinfo{author}{Boute, R.} (\bibinfo{year}{2020}).
\newblock \bibinfo{title}{Use of proximal policy optimization for the joint replenishment problem}.
\newblock {\it \bibinfo{journal}{Computers in Industry}\/},  {\it \bibinfo{volume}{119}\/}, \bibinfo{pages}{103239}.
\bibitem[{Wang \& Lin(2021)}]{wang2021spare}
\bibinfo{author}{Wang, F.}, \& \bibinfo{author}{Lin, L.} (\bibinfo{year}{2021}).
\newblock \bibinfo{title}{Spare parts supply chain network modeling based on a novel scale-free network and replenishment path optimization with q learning}.
\newblock {\it \bibinfo{journal}{Computers \& Industrial Engineering}\/},  {\it \bibinfo{volume}{157}\/}, \bibinfo{pages}{107312}.
\bibitem[{Wang et~al.(2023)Wang, Wang \& Sobey}]{wang2023agent}
\bibinfo{author}{Wang, W.}, \bibinfo{author}{Wang, H.}, \& \bibinfo{author}{Sobey, A.~J.} (\bibinfo{year}{2023}).
\newblock \bibinfo{title}{Agent based modelling for continuously varying supply chains}.
\newblock {\it \bibinfo{journal}{arXiv preprint arXiv:2312.15502}\/}, .
\bibitem[{Yang et~al.(2023)Yang, Liu, Jiang, Zhang, Zhao, Song \& Bian}]{yang2023versatile}
\bibinfo{author}{Yang, X.}, \bibinfo{author}{Liu, Z.}, \bibinfo{author}{Jiang, W.}, \bibinfo{author}{Zhang, C.}, \bibinfo{author}{Zhao, L.}, \bibinfo{author}{Song, L.}, \& \bibinfo{author}{Bian, J.} (\bibinfo{year}{2023}).
\newblock \bibinfo{title}{A versatile multi-agent reinforcement learning benchmark for inventory management}.
\newblock {\it \bibinfo{journal}{arXiv preprint arXiv:2306.07542}\/}, .
\bibitem[{Yavuz \& Kaya(2024)}]{yavuz2024deep}
\bibinfo{author}{Yavuz, T.}, \& \bibinfo{author}{Kaya, O.} (\bibinfo{year}{2024}).
\newblock \bibinfo{title}{Deep reinforcement learning algorithms for dynamic pricing and inventory management of perishable products}.
\newblock {\it \bibinfo{journal}{Applied Soft Computing}\/},  (p. \bibinfo{pages}{111864}).
\bibitem[{Yu et~al.(2020)Yu, Zhou \& Zhang}]{yu2020multi}
\bibinfo{author}{Yu, C.}, \bibinfo{author}{Zhou, Y.}, \& \bibinfo{author}{Zhang, Z.} (\bibinfo{year}{2020}).
\newblock \bibinfo{title}{Multi-agent reinforcement learning for dynamic spare parts inventory control}.
\newblock In {\it \bibinfo{booktitle}{2020 Global Reliability and Prognostics and Health Management (PHM-Shanghai)}\/} (pp. \bibinfo{pages}{1--6}).
\newblock \bibinfo{organization}{IEEE}.

\end{thebibliography}
\newpage 
\appendix

\begin{table}[!htb]
\centering					
\caption{Heterogeneous and Homogeneous PPO best optimal value hyperparameters used when training the agents.}				
\label{Tab:PPO Hyperpara}				
\begin{tabular}{lll}
\toprule				
\textit{Hyperparameters} & \textit{Heterogeneous agent} & \textit{Homogeneous agent} \\ \midrule			

\texttt{fcnet\_hiddens}           & [256,256]                                      & [256,256]                                     \\			
\texttt{preprocessor\_pref}       & deepmind                                       & deepmind                                      \\			
\texttt{placement\_strategy}      & 'PACK'                                         & -                                             \\			
\texttt{vf\_loss\_coeff}          & 1.0                                            & 1.0                                           \\			
\texttt{lstm\_cell\_size}         & 256                                            & 256                                           \\			
\texttt{sgd\_minibatch\_size}     & 128                                            & 512                                           \\			
\texttt{Learning rate lr}         & 5e-05                                          & 0.0001                                        \\			
\texttt{vf\_share\_layers}        & -1                                             & -1                                            \\			
\texttt{Discount factor (gamma)}  & 0.99                                           & 0.99                                          \\			
\texttt{train\_batch\_size}       & 4000                                           & 4000                                          \\			
\texttt{attention\_dim}           & 64                                             & -                                             \\			
\texttt{clip\_param}              & 0.3                                            & 0.3                                           \\			
\texttt{kl\_target}               & 0.01                                           & 0.01                                          \\			
\texttt{max\_seq\_len}            & 20                                             & 20                                            \\			
\texttt{fcnet\_activation}        & tanh                                           & tanh                                          \\			
\texttt{conv\_activation}         & relu                                           & relu                                          \\			

\bottomrule			
\end{tabular}				
\end{table} 

\begin{table}[!htb]
\centering					
\caption{Heterogeneous and Homogeneous SAC hyperparameters used when training the agents.}				
\label{Tab:SAC Hyperpara}				
\begin{tabular}{lll}
\toprule				
\textit{Hyperparameters} & \textit{Heterogeneous agent} & \textit{Homogeneous agent} \\ \midrule			

Min history to start learning & 80K frames & - \\			
prioritized\_replay\_eps & 1e-06 & 1e-06 \\			
max\_seq\_len & 20 & 20 \\			
prioritized\_replay\_alpha & 0.6 & 0.6 \\			
Multi-step returns n & 3 & - \\			
Exploration $\gamma$ & 0.0 & - \\			
Adam $\epsilon$ & $1.5 \times 10^{-4}$ & - \\			
Noisy Nets $\rho_0$ & 0.5 & - \\			
Adam learning rate & 0.0000625 & - \\			
evaluation\_sample\_timeout & 180.0 & 120.0 \\			
prioritized\_replay\_beta & 0.4 & 0.4 \\			
tau & 0.005 & 0.005 \\			
fcnet\_activation & relu & relu \\			
Distributional atoms & 51 & - \\			
mean\_dim & 84 & 84 \\			
learning rate & 0.001 & 0.001 \\			
actor\_learning\_rate & 0.003 & 0.0003 \\			
critic\_learning\_rate & 0.0003 & 0.0003 \\			
entropy\_learning\_rate & 0.0003 & 0.0003 \\
gamma& 0.99 & 0.99 \\

\bottomrule			
\end{tabular}				
\end{table}

\end{document}